\documentclass[useAMS,usenatbib]{mn2e}
\usepackage{graphicx}
\usepackage{appendix}
\usepackage{lscape}
\usepackage{longtable}
\usepackage{subfigure}

\def\lsim{\,\lower2truept\hbox{${<\atop\hbox{\raise4truept\hbox{$\sim$}}}$}\,}
\def\gsim{\,\lower2truept\hbox{${> \atop\hbox{\raise4truept\hbox{$\sim$}}}$}\,}

\title[H-ATLAS: the lensed galaxy sample]
  {The {\it Herschel}\thanks{{\it Herschel} is an ESA space observatory
      with science instruments provided by European-led Principal
      Investigator consortia and with important participation from
      NASA.}-ATLAS: a sample of 500$\mu$m-selected lensed
    galaxies over 600 square degrees}
\author[M. Negrello et al.]
{M.~Negrello,$^{1}$\thanks{NegrelloM@cardiff.ac.uk},
S.~Amber$^{2}$,
A.~Amvrosiadis$^{1}$,
Z.-Y. Cai$^{3}$,
A.~Lapi$^{4,5}$, 
J.~Gonzalez-Nuevo$^{6}$, 
\newauthor
G.~De Zotti$^{7,5}$,
C.~Furlanetto$^{8,9}$,
S.~Maddox$^{1,10}$,
M.~Allen$^{1}$, 
T.~Bakx$^{1}$,
R.~S.~Bussmann$^{11}$, 
\newauthor
A.~Cooray$^{12}$,
G.~Covone$^{13}$
L.~Danese$^{5}$, 
H.~Dannerbauer$^{14}$,
H.~Fu$^{15}$,
J.~Greenslade$^{16}$,
\newauthor
M.~Gurwell$^{17}$,
R.~Hopwood$^{16}$,
L.~V.~E.~Koopmans$^{18}$,
N.~Napolitano$^{19}$,
H.~Nayyeri$^{12}$,
\newauthor
A.~Omont$^{20,21}$,
C.~E.~Petrillo$^{18}$,
D.~A.~Riechers$^{11}$,
S.~Serjeant$^{2}$, 
C.~Tortora$^{18}$,
E.~Valiante$^{1}$,
\newauthor
G.~Verdoes~Kleijn$^{18}$, 
G.~Vernardos$^{18}$,
J.~L.~Wardlow$^{22}$,
M.~Baes$^{23}$,
A.~J.~Baker$^{24}$,
\newauthor
N.~Bourne$^{10}$,
D.~Clements$^{16}$,
S.~M.~Crawford$^{25}$,
S.~Dye$^{8}$,
L.~Dunne$^{1,10}$,
S.~Eales$^{1}$,
\newauthor
R.~Ivison$^{26,10}$,
L.~Marchetti$^{2,25}$,
M.~J.~Micha{\l}owski$^{10}$,
M.~W.~L.~Smith$^{1}$, 
M.~Vaccari$^{27,28}$, 
\newauthor
P.~van der Werf$^{29}$ \\
$~$ \\
{\it Affiliations are listed at the end of the paper}
}

\pagerange{\pageref{firstpage}--\pageref{lastpage}} \pubyear{2002}

\def\LaTeX{L\kern-.36em\raise.3ex\hbox{a}\kern-.15em
    T\kern-.1667em\lower.7ex\hbox{E}\kern-.125emX}

\begin{document}

\label{firstpage}

\maketitle

\begin{abstract}
We present a sample of 80 candidate strongly lensed galaxies with flux
density above 100\,mJy at 500\,$\mu$m extracted from the 
{\it Herschel} Astrophysical Terahertz Large Area Survey ({\it H}-ATLAS), over an
area of 600\,deg$^{2}$. 
Available imaging and spectroscopic data allow us to confirm the
strong lensing in 20 cases and to reject it in one case. For other 8
objects the lensing scenario is strongly supported by the presence of two sources along
the same line of sight with distinct photometric redshifts. 
The remaining objects await more follow-up observations to confirm
their nature. The lenses and the background sources have
  median redshifts $z_{\rm L}=0.6$ and $z_{\rm S}=2.5$,
  respectively, and are observed out to $z_{\rm L}=1.2$ and $z_{\rm S}=4.2$. We measure the number counts of candidate lensed
galaxies at 500\,$\mu$m and compare them with theoretical predictions,
finding a good agreement for a maximum magnification of the background
sources in the range 10-20. These values are consistent with the
magnification factors derived from the lens modeling of individual
systems. The catalogue
presented here provides sub-mm bright targets for follow-up
observations aimed at exploiting gravitational lensing to study with
un-precedented details the morphological and 
dynamical properties of dusty starforming regions in $z\gsim1.5$ galaxies.
\end{abstract}


\section{Introduction}

Wide area extragalactic surveys performed at sub-millimeter (sub-mm) to
millimeter (mm) wavelengths with the {\it Herschel} Space Observatory
\citep{Pil10} and the South Pole Telescope \citep[SPT;][]{Carlstrom11} have led to
the discovery of several dusty star forming galaxies (DSFGs) at
$z\gsim1$ whose luminosity is magnified by a foreground galaxy or a
group/cluster of galaxies
\citep{Neg10,Neg14,Cox11,Con11,Buss12,Buss13,Vie13,Wardlow13,Messias14,Calanog14,Dye15,Nayyeri16,Spilker16}. The selection of strongly lensed
galaxies at these wavelengths is made possible by the (predicted) steep number counts
of high redshift sub-mm galaxies \citep{Blain96,Neg07}; in fact almost exclusively those
galaxies whose flux density has been boosted by an event of lensing can be
observed above a certain threshold, namely $\sim100\,$mJy at
500\,$\mu$m \citep{Neg10,Wardlow13}. 

By means of a simple selection in flux density at 500\,$\mu$m, \cite{Neg10}
produced the first sample of 5 strongly lensed galaxies from
16\,deg$^{2}$ of the sky observed with {\it Herschel} during the Science Demonstration Phase (SDP)
as part of the {\it Herschel} Astrophysical
Terahertz Large Area Survey\footnote{www.h-atlas.org} \citep[{\it
  H}-ATLAS hereafter]{E10}. 
Preliminary source catalogues derived from the full {\it H}-ATLAS were then used to identify the sub-mm brightest candidate lensed
galaxies for follow-up observations with both ground based
and space telescopes to measure their redshifts \citep{Frayer11,Val11,Harris12,Lupu12,Om11,Om13,George13,Messias14} and
confirm their nature \citep{Neg10,Neg14,Fu12,Buss12,Buss13,Calanog14}. 
Using the same methodology, i.e. a cut in flux density at 500\,$\mu$m, \cite{Wardlow13} have
identified 11 lensed galaxies over 95\,deg$^{2}$ of the {\it
  Herschel} Multi-tiered Extragalactic
Survey \citep[HerMES;][]{Oliver12}, while, more recently, \cite{Nayyeri16} have
published a catalogue of 77 candidate lensed galaxies with $F_{\rm 500}\ge100\,$mJy extracted from
the HerMES Large Mode Survey \citep[HeLMS;][]{Oliver12}  and the {\it Herschel} Stripe 82 Survey
\citep[HerS;][]{Viero14}, over an area of 372\,deg$^{2}$. 
All together the extragalactic surveys carried out with {\it Herschel} are expected to deliver a sample of 
more than a hundred of sub-mm bright strongly lensed galaxies, which,
as argued by \cite{Gonz12},  might increase 
to over a thousand if the selection is based on the steepness of the 
luminosity function of DSFGs \citep{Lapi11} rather than that of the
number counts.
Similarly, at millimeter wavelengths, the SPT survey has already
discovered severals tens of lensed galaxies
\citep[e.g.][]{Vie13,Spilker16} and other lensing events have been
found in the {\it Planck} all sky surveys \citep{PlanckXXVII,Canameras15,Harrington16}

Having a large sample of strongly lensed DSFGs is important for several reasons: (i) thanks to the boosting in luminosity 
and increase in angular size offered by gravitational lensing, distant galaxies can be studied with unprecedented details down to sub-kpc scales.
For sub-mm/mm selected sources this means understanding the morphological and dynamical properties of individual 
giant molecular clouds in a statistically significant sample of dusty galaxies across the peak of the cosmic star formation history 
of the Universe \citep[e.g.][]{Swin10,Dye15}; (ii) the observed lensed morphology is determined by the content and the spatial distribution of the total (baryonic+dark) matter 
in the foreground galaxy; therefore, by means of high resolution imaging data at sub-mm/mm wavelengths $-$ as now provided by  the
Atacama Large Millimeter Array (ALMA) $-$ gravitational lensing allows the detection of low-mass substructures in the lenses, whose abundance can be used to test the
cold dark matter scenario of structure formation on small scales \citep[e.g.][]{MS98,MM01,DK02,Vegetti12,Hezaveh13,Inoue16,Hezaveh16}; 
in this respect, it is worth noticing that for sub-mm/mm selected lensed galaxies the contamination of the lens $-$ typically a passively evolving elliptical galaxy $-$ 
to the sub-mm emission of the background galaxy is negligible. Therefore, the modeling of the lensed morphology does not suffer from uncertainties on the lens subtraction, 
as usually happens  in optically selected lensing systems; furthermore, since the sample of lensed galaxies is constructed by exploiting the properties of the background galaxy alone, 
i.e. its sub-mm flux density, it is less biased against the redshift
and the mass of the lens, compared to standard optical/spectroscopic 
selection techniques, thus allowing us to probe the 
mass distribution of foreground galaxies out to $z\gsim1$\citep[e.g.][]{Dye14}; (iii) the statistics of lenses, as well as the distribution and extent of the image separations, 
depend on the dark-matter and dark-energy content of the Universe; therefore large samples of gravitational lenses can be used to 
constrain cosmological parameters \citep[e.g.][]{Grillo2008,Eales15}.

Here we present the sample of candidate lensed galaxies
with $F_{\rm 500}\ge100\,$mJy extracted from the $\sim$600\,deg$^{2}$
of the full {\it
  H}-ATLAS. 
The work is organized as follows. The {\it H}-ATLAS catalogue and available
ancillary data sets are presented
in Sec.\,\ref{sec:data}. In Sec.\,\ref{sec:lens_selection} we describe the
selection of the candidate lensed galaxy and discuss the properties
of the sample. In Sec.\,\ref{sec:number_counts} we derive the number counts of
lensed galaxies and compare them with model predictions. Conclusions
are summarized in Sec.\,\ref{sec:conclusions}. \\
The modeling of the number counts is made assuming a spatially flat cosmological model with present-day matter density in units of the critical density, 
$\Omega_{0,m}=0.308$ and Hubble constant
$H_0=67.7\,\hbox{km}\,\hbox{s}^{-1}\,\hbox{Mpc}^{-1}$, spectrum of primordial density perturbations with slope $n = 1$ and 
normalization on a scale of $8\,h^{-1}$\,Mpc $\sigma_{8} = 0.81$ \citep{Planck_parameters2015}.

\begin{table*}
\begin{center}
\caption{Blazars with $F_{\rm 500}\ge100\,$mJy identified in the {\it
    H}-ATLAS fields. {\it Herschel}/SPIRE flux densities are
  provided together with 1.4\,GHz flux densities and spectroscopic
  redshifts (from NED).}\label{tab:blazars}
\vspace{-0.0cm}
\begin{tabular}{llrrrrr}
\hline 
\hline 
\multicolumn{1}{c}{{\it H}-ATLAS IAU name} &
\multicolumn{1}{c}{NED name} & 
\multicolumn{1}{c}{$F_{\rm 250}$}  &
\multicolumn{1}{c}{$F_{\rm 350}$}  &
\multicolumn{1}{c}{$F_{\rm 500}$}  &
\multicolumn{1}{c}{$F_{\rm 1.4GHz}$} &
\multicolumn{1}{c}{$z_{\rm spec}$} \\
 &
 & 
\multicolumn{1}{c}{(mJy)} &
\multicolumn{1}{c}{(mJy)} &
\multicolumn{1}{c}{(mJy)} &
\multicolumn{1}{c}{(mJy)} &
\\
\hline
HATLASJ090910.1+012134$^{\dagger,\ddagger}$  &  [HB89] 0906+015                   & 256.5$\pm$6.4    &  327.1$\pm$7.4  &  375.3$\pm$8.1    &  760$\pm$23   & 1.024905  \\ 
HATLASJ121758.7$-$002946                            &  PKS 1215$-$002                    & 122.7$\pm$7.4    &  152.3$\pm$8.1  &   177.5$\pm$8.4   &  452$\pm$16   &  0.418406  \\
HATLASJ141004.7+020306$^{\ddagger}$             &  [HB89] 1407+022                   & 119.4$\pm$7.3   &  151.0$\pm$8.4   &   176.0$\pm$8.7   & 334$\pm$10    &  1.253000 \\
HATLASJ131028.7+322043                                &	[HB89] 1308+326                  &  259.1$\pm$6.7  &  363.1$\pm$7.7   &  452.2$\pm$8.1    &  1687$\pm$51  &  0.998007 \\
HATLASJ125757.3+322930                                &   7C 1255+3245                      &  143.7$\pm$7.3  &  188.4$\pm$8.2   &   214.9$\pm$8.6    &  653$\pm$20  &  0.805949 \\
HATLASJ133307.3+272518                                &   87GB 133047.7+274044       &   89.3$\pm$7.4   &  104.6$\pm$8.1   &   117.1$\pm$8.3    &  218$\pm$7     &  2.126000 \\
HATLASJ131736.4+342518                                &   [HB89] 1315+346                  &  77.1$\pm$7.2    &   99.5$\pm$8.0    &    112.0$\pm$8.7   &  529$\pm$16   &  1.055411 \\
HATLASJ014503.3$-$273332                            &   [HB89] 0142$-$278               &  131.5$\pm$5.7  &   179.1$\pm$6.3   &   233.5$\pm$7.2   &  923$\pm$28   &  1.155000 \\
HATLASJ224838.6$-$323550                            &   [HB89] 2245$-$328               &  119.2$\pm$5.5  &   152.8$\pm$5.8   &   194.8$\pm$6.7   &  708$\pm$21  &  2.268000 \\
HATLASJ014309.9$-$320056                            &   PKS 0140$-$322                    &   96.0$\pm$5.3   &   119.5$\pm$5.9   &   122.4$\pm$7.2   &  76$\pm$2      &  0.375100 \\
HATLASJ235347.4$-$303745                            &   PKS 2351$-$309                    &  77.1$\pm$5.1    &    96.6$\pm$5.8    &   103.1$\pm$7.0   &  398$\pm$14  &   - \\
\hline
\hline
\multicolumn{5}{l}{$^{\dagger}$ also reported in \cite{Gonz10}.} \\
\multicolumn{5}{l}{$^{\ddagger}$ also reported in \cite{LopezCaniego13}.} \\
\end{tabular} 
\end{center}
\end{table*}

\begin{table*}
\begin{center}
\caption{Dusty stars with $F_{\rm 500}\ge100\,$mJy identified in the {\it
    H}-ATLAS fields. The {\it Herschel}/SPIRE flux densities are
  provided.}\label{tab:dusty_stars}
\vspace{-0.0cm}
\begin{tabular}{llrrr}
\hline 
\hline 
\multicolumn{1}{c}{{\it H}-ATLAS IAU name} &
\multicolumn{1}{c}{NED name} & 
\multicolumn{1}{c}{$F_{\rm 250}$}  &
\multicolumn{1}{c}{$F_{\rm 350}$}  &
\multicolumn{1}{c}{$F_{\rm 500}$}   \\
 &
 & 
\multicolumn{1}{c}{(mJy)} &
\multicolumn{1}{c}{(mJy)} &
\multicolumn{1}{c}{(mJy)} \\
\hline
HATLASJ132301.6+341649          &  SDSS J132301.74+341647.8   & 124.2$\pm$7.3    &  144.6$\pm$8.2  &  137.0$\pm$8.7  \\ 
HATLASJ225739.6$-$293730      &   $\alpha$PsA (Fomalhaut)                &   825.2$\pm$5.7   &   433.0$\pm$6.1    &   292.1$\pm$7.3   \\             
HATLASJ012657.9$-$323234      &   R Sclulptoris                                       &   795.6$\pm$5.5    &    541.7$\pm$6.0   &   242.4$\pm$7.1   \\ 
\hline
\hline
\end{tabular} 
\end{center}
\end{table*}

\section{Data sets}\label{sec:data}

 {\it H}-ATLAS is the widest-area extragalactic survey
undertaken with {\it Herschel}, imaging around
600\,deg$^2$ of the sky in five far-infrared (far-IR) to sub-mm bands,
100, 160, 250, 350 and 500\,$\mu$m, using the PACS \citep{Pog10}
and SPIRE \citep{Gri10} instruments, in parallel mode.
{\it H}-ATLAS covers 5 fields: three fields on the celestial equator,
each about 54\,deg$^{2}$ in size and approximatively located at right
ascension RA = 9h, 12h and 15h, 
a large ($\sim$170\,deg$^{2}$) field close to the North Galactic Pole
(NGP) and an even larger field ($\sim$270\,deg$^{2}$)
near the South Galactic Pole (SGP).

The fields were selected to minimize bright continuum emission from
dust in the Galaxy, as traced at 100\,$\mu$m by the Infrared Astronomical Satellite (IRAS), and
to benefit from existing data at other wavelengths, particularly
spectroscopic optical data provided by other major surveys of the
nearby Universe, e.g. the Galaxy and Mass Assembly (GAMA)
survey\footnote{The {\it H}-ATLAS equatorial fields are also referred
  to as the {\it H}-ATLAS/GAMA fields}
\citep{Driver09,Driver16}, the Sloan Digital Sky Survey \citep[SDSS;][]{Aba09}, the
2-Degree-Field Galaxy Redshift Survey
\citep[2dFGRS;][]{Colless2001} and the Kilo-Degree Survey
\citep[KiDS;][]{deJong2015}. Apart from optical imaging and spectroscopy, 
the fields have imaging data at near-infrared wavelengths from the UK Infrared
Deep Sky Survey Large Area Survey \citep[UKIDSS-LAS;][]{Lawrence2007}
and the VISTA Kilo-Degree Infared Galaxy Survey
\citep[VIKING;][]{Edge2013}.
Radio-imaging data in the fields are provided by the the Faint Images
of the Radio Sky at Twenty-cm (FIRST) survey and the NRAO Very Large Array Sky Survey (NVSS).

Candidate lensed galaxies were selected from the {\it Herschel}/SPIRE
\citep{Gri10} catalogue of the  {\it H}-ATLAS Data Release 1 and 2
\citep{Val16}.
These catalogues are created in two stages. First, the Multi-band
Algorithm for source Detection and eXtraction (MADX, Maddox et
al. in prep.) 
is used to identify the 2.5$\,\sigma$ peaks in the 250\,$\mu$m
maps and to measure the flux densities at the position of
those peaks in all the SPIRE bands. 
The maps used for the flux measurements have been filtered with a
matched-filter 
technique \citep{Chapin11} to minimize instrumental and confusion
noise. 
Second, only sources with signal-to-noise ratio $\ge4$ in at least one of the three SPIRE bands are kept in the final catalogue.
The $4\sigma$ detection limit at 250$\,\mu$m for a point source ranges 
from 24\,mJy in the deepest regions of the maps (where tiles overlap)
to 29\,mJy in the non-overlapping regions.
For more details about the source extraction procedure and the flux density measurements
we refer the reader to \cite{Val16}. The catalogue and the maps will
be made available at the {\it H}-ATLAS website: www.h-atlas.org.

Flux density measurements for extended galaxies are currently only available for sources
detected in the {\it H}-ATLAS equatorial fields. Therefore, in the
present analysis, we only use point source flux densities. This is not
an issue for lensed galaxies as they do appear as point sources in the
SPIRE
maps. Extended galaxies are treated as contaminants to be removed
from the final sample and they are identified by means of available
optical imaging data; therefore their precise flux density at sub-mm
wavelengths is not a concern. We warn the reader against using
the number of local galaxies with $F_{\rm 500}\geq100\,$mJy derived here for
comparison with theoretical models, as such number is significantly
underestimating the true density of local galaxies that have 500\,$\mu$m
flux density above the adopted threshold.

\begin{figure}
  \hspace{-0.3cm}\includegraphics[width=8.7cm]{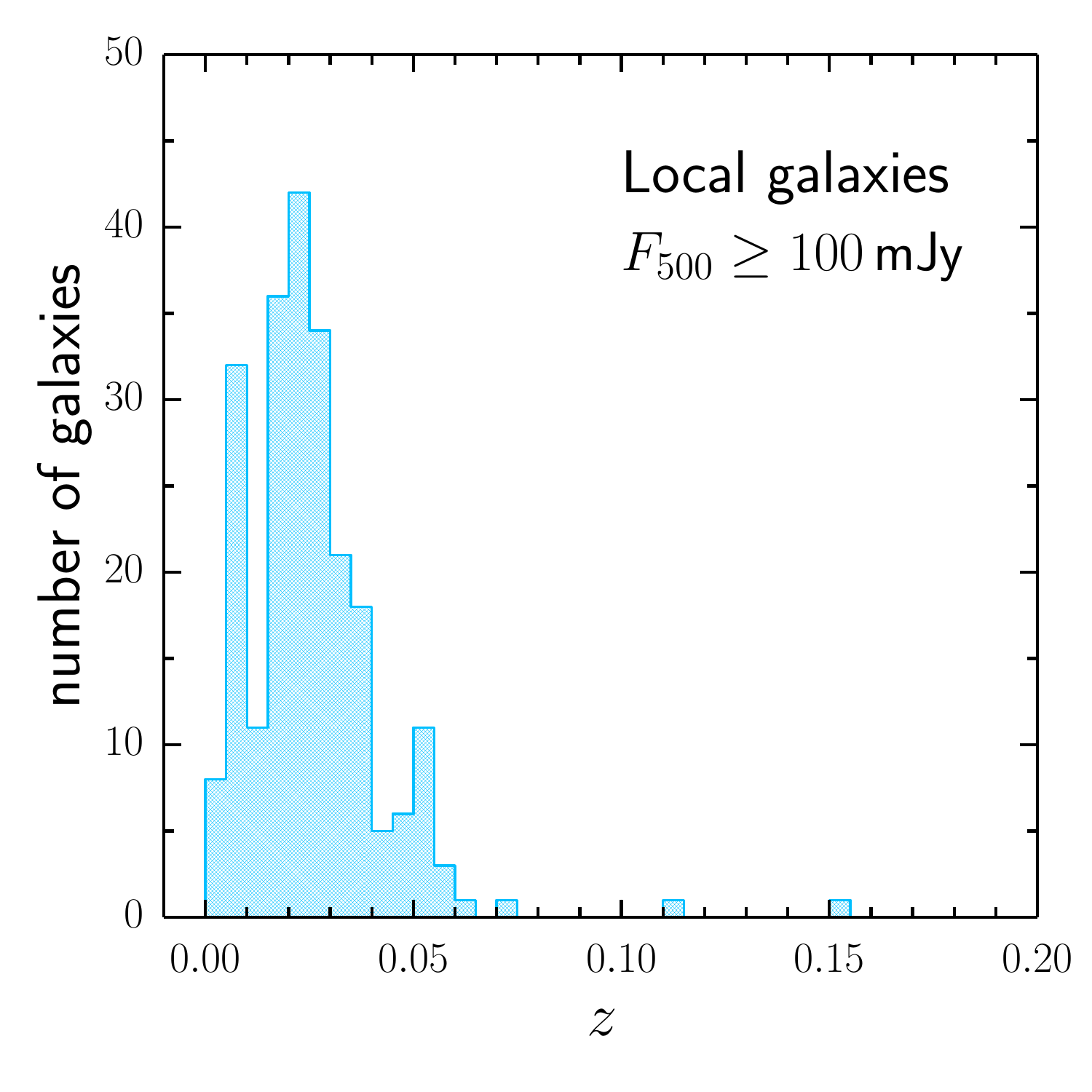}
 \vspace{-0.7cm}
 \caption{Redshift distribution of local galaxies with F$_{\rm
     500}\ge100$\,mJy (blue)  identified in the {\it H}-ATLAS
   equatorial fields. The sources at $z>0.1$ are HATLASJ090734.8+012504 ($z=0.102315$)
and HATLASJ120226.7$-$012915 ($z=0.150694$).} 
 \label{fig:Nz_local}
\end{figure}

\begin{figure*}
  \hspace{0.0cm}\vspace{0.0cm}\includegraphics[width=18.0cm]{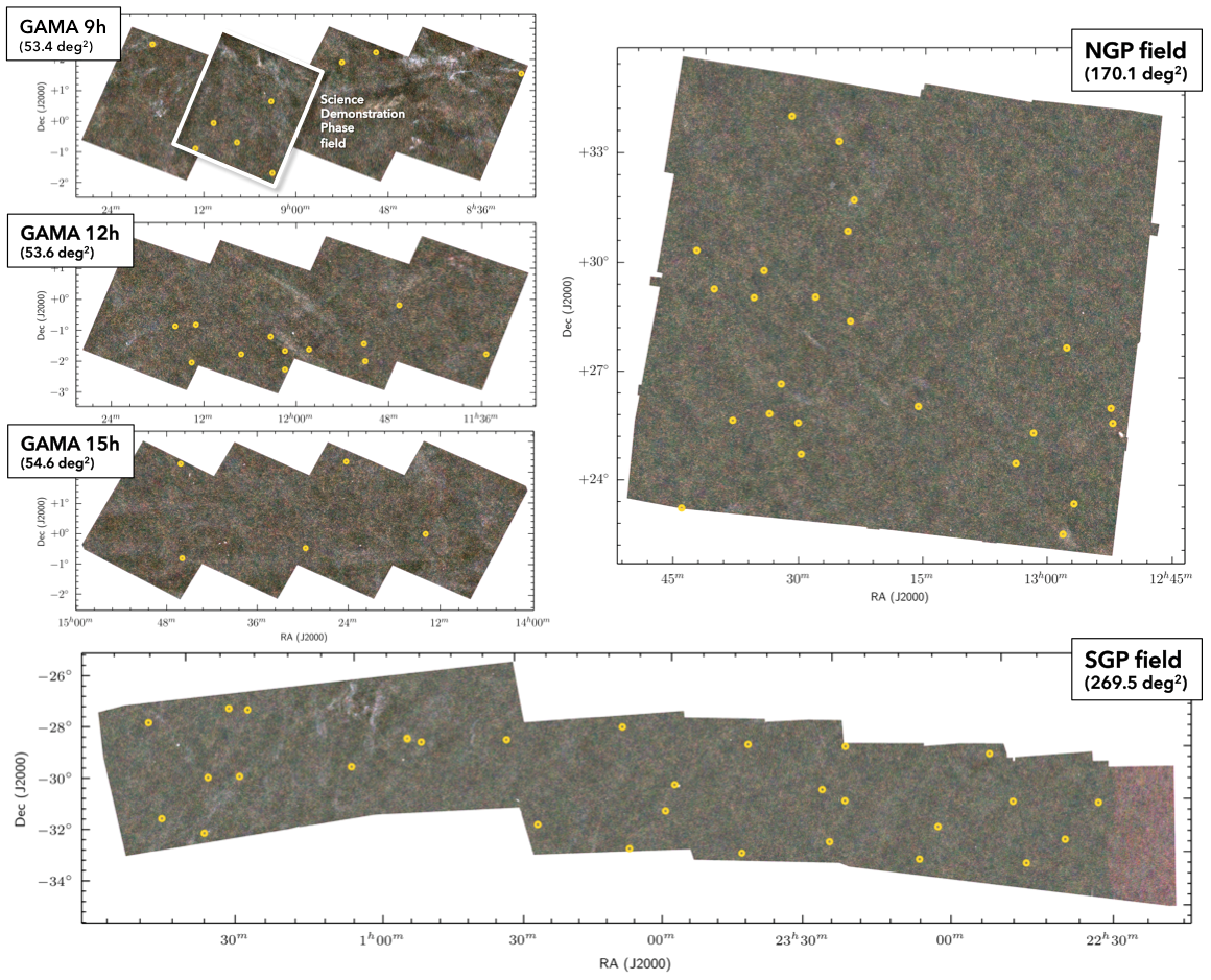} 
\vspace{0.0cm}\caption{{\it Herschel}/SPIRE color maps of the {\it
    H}-ATLAS fields. The yellow circles mark the 
position of the 80 candidate lensed galaxies 
with $F_{\rm 500}\ge100\,$mJy. 
}
\label{fig:HATLAS_fields}
\end{figure*}

The redshift of several lensed galaxies discovered in {\it H}-ATLAS has been
constrained through the detection of rest-frame far-infrared/sub-mm emission lines from carbon monoxide (CO), 
water vapor (H$_{\rm 2}$O), ionized carbon (CII) and doubly ionized oxygen ([OIII]), using Z-spec
\citep{Lupu12}, GBT/Zspectrometer \citep{Frayer11,Harris12}, PdBI
\citep{Neg10,Om11,Om13,Yang2016,George14}, CARMA (Riechers et al. in
prep.),  and the {\it Herschel}/SPIRE Fourier Transform
Spectrometer \citep{Val11,George13}. 

Optical spectra of the lenses in  confirmed {\it H}-ATLAS
selected lensing systems  were taken with the William
Herschel Telescope (WHT), the Apache
Point Observatory 3.5-meter telescope \citep{Neg10}, the New-Technology
Telescope \citep[NTT;][]{Amber2015}, the Gemini-South telescope
\citep{Buss13} and the Anglo Australian Telescope as part of GAMA. A
project is ongoing to get optical spectra with the South African Large
Telescope \citep[SALT;][Marchetti et al. in prep.]{Serjeant2016}. 

Near infrared follow-up observations with the Hubble Space Telescope
({\it HST}) and the {\it Keck} telescope in Adaptive Optics (AO) are
available for tens of candidate lensed galaxies in the
{\it H}-ATLAS fields \citep{Fu12,Neg14,Buss13,Calanog14}. Some of these sources have
also been
observed with the Sub-millimeter Array at $\sim$0.5$^{\prime\prime}$ resolution at
880$\,\mu$m \citep{Neg10,Buss12,Buss13}, while six of them have been recently imaged
in band 7 with the Atacama Large Millimeter Array (ALMA) with better
resolution and sensitivity (PI: Eales). A lens modeling of the ALMA data
will be presented in future papers (Dye et al in
prep.; Negrello et al. in prep.).

\begin{table}
  \begin{center}
    \caption{Number of sources with $F_{\rm 500}\ge100\,$mJy classified as candidate lensed
      galaxies, local (late-type) galaxies, blazars and dusty stars in
  the {\it H}-ATLAS fields.}\label{tab:number_of_sources}
    \vspace{-0.0cm}
    \begin{tabular}{lcccc}
      \hline 
      \hline 
      & candidate           & local       & blazars & dusty \\
      & lensed galaxies  & galaxies &              & stars     \\
\hline
\hline
      9h\,field    &  9    &  10    & 1 & 0 \\
      12h\,field  &  12  &  18    & 1 & 0 \\
      15h\,field  &  5    &  32    & 1 & 0 \\
      NGP field   &  24  &  84    &  4 & 1 \\
      SGP field    &  30  &  87    &  4 & 2\\
\hline
      Total         & 80   & 231  & 11 & 3\\
\hline
\hline
    \end{tabular} 
  \end{center}
\end{table}

\section{Candidate lensed galaxies}\label{sec:lens_selection}

We describe below the procedure adopted to identify
candidate lensed galaxies in the {\it H}-ATLAS and discuss the main properties of the
sample.

\subsection{Selection}

According to model predictions \citep{Blain96,P02,P03,Lapi06,Lapi11,Neg07,Neg10}, the
number counts of un-lensed DSFGs are expected to drop abruptly at 500$\,\mu$m flux
densities of $\simeq$100$\,$mJy, as an effect of the intrinsically
steep luminosity function and high redshift ($z\gsim1.5$) of these sources. Therefore, we start
by selecting all the sources with $F_{\rm 500}\ge100\,$mJy. They
are 325 in total. A fraction of these sources are either low-redshift
($z\lsim0.1$) spiral galaxies or flat spectrum radio sources
\citep[e.g.][]{Neg07}. Both of these classes of objects are considered as
``contaminants'' for the purpose of this paper, and therefore need to
be identified and removed. We used the  interactive software sky atlas
{\sc Aladin} \citep{Aladin00} to inspect available optical, infrared
and radio imaging data and to query the
NASA/IPAC Extragalactic
Database\footnote{http://ned.ipac.caltech.edu/} (NED) around the position of
each source. 

We identified 11 blazars, according to their intense radio emission ($F_{\rm
  1.4GHz}>100\,$mJy) and the rising/flat SED from
sub-millimeter to radio wavelengths. They are listed in Table\,\ref{tab:blazars}. Their number density is consistent with theoretical expectations \citep[e.g.][]{Tucci11}. 
Two of them were previosuly reported by \cite{LopezCaniego13} 
who searched for blazars in 135\,deg$^{2}$ of the {\it H}-ATLAS equatorial fields. 

\begin{figure}
  \hspace{0.0cm}
  \begin{minipage}[b]{0.51\linewidth}
    \centering \resizebox{2.0\hsize}{!}{
      \hspace{-1.0cm}\includegraphics{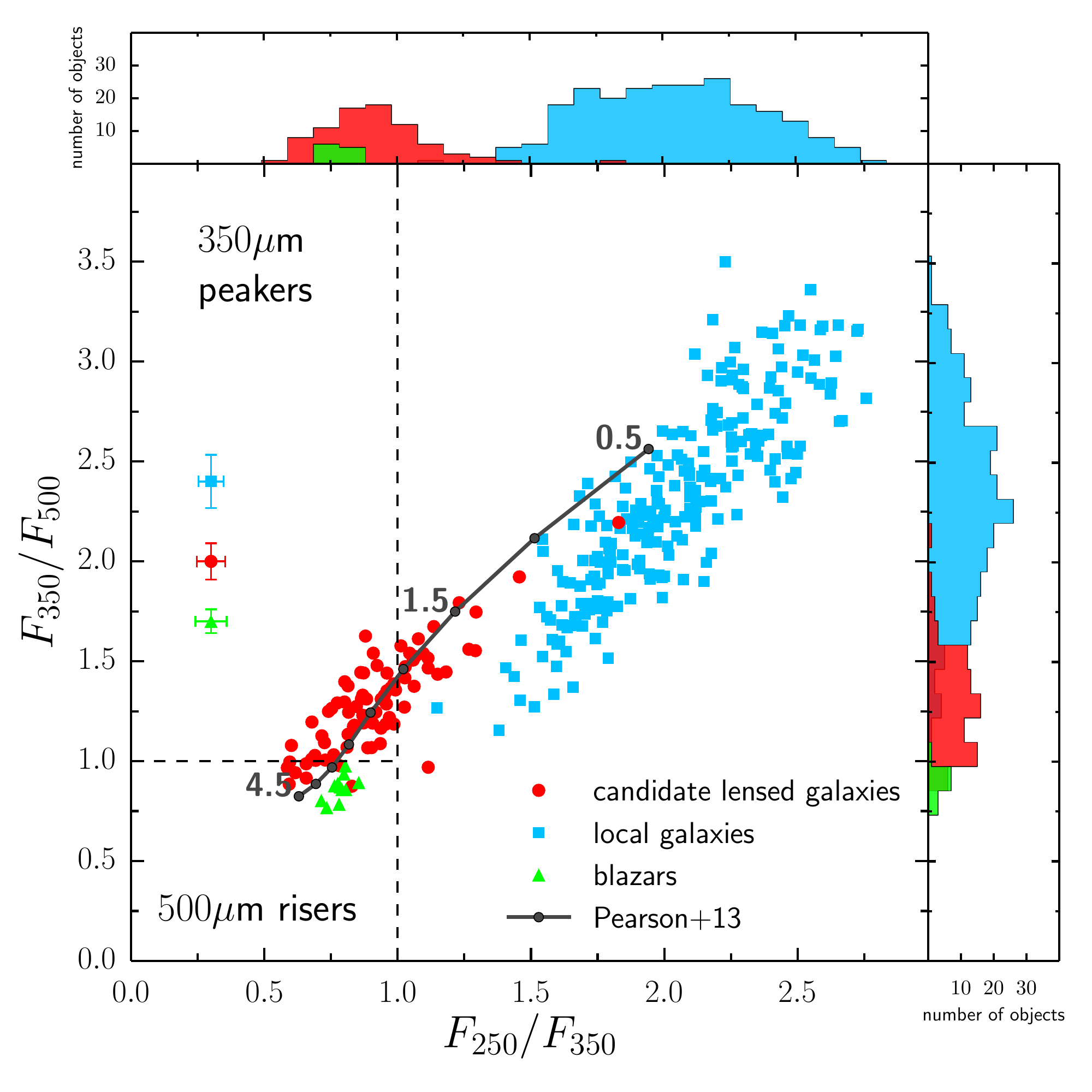}} 
  \end{minipage}
 \vspace{-0.3cm}
 \caption{{\it Herschel}/SPIRE colour-colour diagram of candidate lensed galaxies
(red dots), local galaxies (blue squares) and blazars (green triangles) with
   F$_{\rm500}\ge100$\,mJy identified in the {\it H}-ATLAS fields. The histograms show the number of the source as a function of flux density ratio. 
Typical error bars for the different source populations are shown on the left. The black line is the track of
   the Pearson et al. (2013) empirical template for redshifts in the range
   [0.5,4.5] in steps of 0.5 (in increasing order from the top-right
   to the bottom-left corner), as marked by the black dots. }
 \label{fig:colorcolor_plot}
\end{figure}

231 spiral-like galaxies with an angular size exceeding several
arcseconds in optical images were categorized as local galaxies.
All were confirmed as local galaxies via available
spectroscopic redshift. Their redshift distribution is shown in
Fig.\,\ref{fig:Nz_local}. Two galaxies have $z>0.1$: HATLASJ090734.8+012504 ($z=0.102315$)
and HATLASJ120226.7$-$012915 ($z=0.150694$), which would place them in the
right redshift range for acting as gravitational lenses. However in
these cases we can exclude that the sub-millimeter emission, as measured
by {\it Herschel}, is coming from a more distant background galaxy
that has been
gravitationally lensed. In fact, in both cases, the {\it Herschel}/SPIRE colors are
consistent with the measured spectroscopic redshift: $F_{\rm
  250}/F_{\rm 350}=2.4$ and $F_{\rm 350}/F_{\rm 500}=3.1$  for
HATLASJ090734.8+012504, and similarly $F_{\rm
  250}/F_{\rm 350}=2.4$ and $F_{\rm 350}/F_{\rm 500}=2.7$ for
HATLASJ120226.7$-$012915 (see also section\,\ref{subsec:colours_redshifts} and the related fig.\,\ref{fig:colorcolor_plot}). 

After this process, we are left with 83 objects, three of which are
found to be dusty stars: HATLASJ132301.6+341649 in the NGP field and
HATLASJ225739.6−293730 and HATLASJ012657.9−323234 in the SGP
field. They are listed in Table\,\ref{tab:dusty_stars}. While the ones in the SGP field are well known stars with infrared emission, the one in the NGP was identified as such 
based on its star-like classification in SDSS, its point like appearance in available HST/WFC3 imaging data and its measured non null proper motion (23.3$\pm$4.5 mas\,yr$^{-1}$) in the United States Naval Observatory (USNO) catalog.
The remaining 80 sources are retained as {\it candidate lensed
  galaxies}. Table\,\ref{tab:number_of_sources} summarizes the
result of the classification. It is worth mentioning that our selection method picked up an extra candidate lensed galaxy, HATLASJ120735.6+005400, in the 12h equatorial field, which, however, is not included 
in the current sample. 
Indeed, this source turned out to be an asteroid (3466 Ritina) and as such it was removed from the publicly released {\it H}-ATLAS catalogue \citep{Val16}.

The 80 candidate lensed galaxies are listed in Table\,\ref{tab:lens_candidates_F500ge100mJy} together with their {\it Herschel}/SPIRE photometry and other additional information. 
Fig.\,\ref{fig:HATLAS_fields} shows their distribution within the five {\it
  H}-ATLAS fields. 
The 16\,deg$^{2}$ SDP field, marked by the white frame in the figure, 
was a lucky choice in terms of detecting lensed galaxies: 
it contains 5 confirmed lensed sources with $F_{\rm 500}\ge100\,$mJy \citep{Neg10}. 
On average, within a similar area, we would have expected $80/602\times16\simeq2$ 
lensed galaxies with 500\,$\mu$m flux density above the 100\,mJy
threshold. 

\subsection{Confirmed lensed and unlensed galaxies}

\begin{figure*}
\centering
  \vspace{0.0cm}\hspace{0.0cm}\includegraphics[width=18cm]{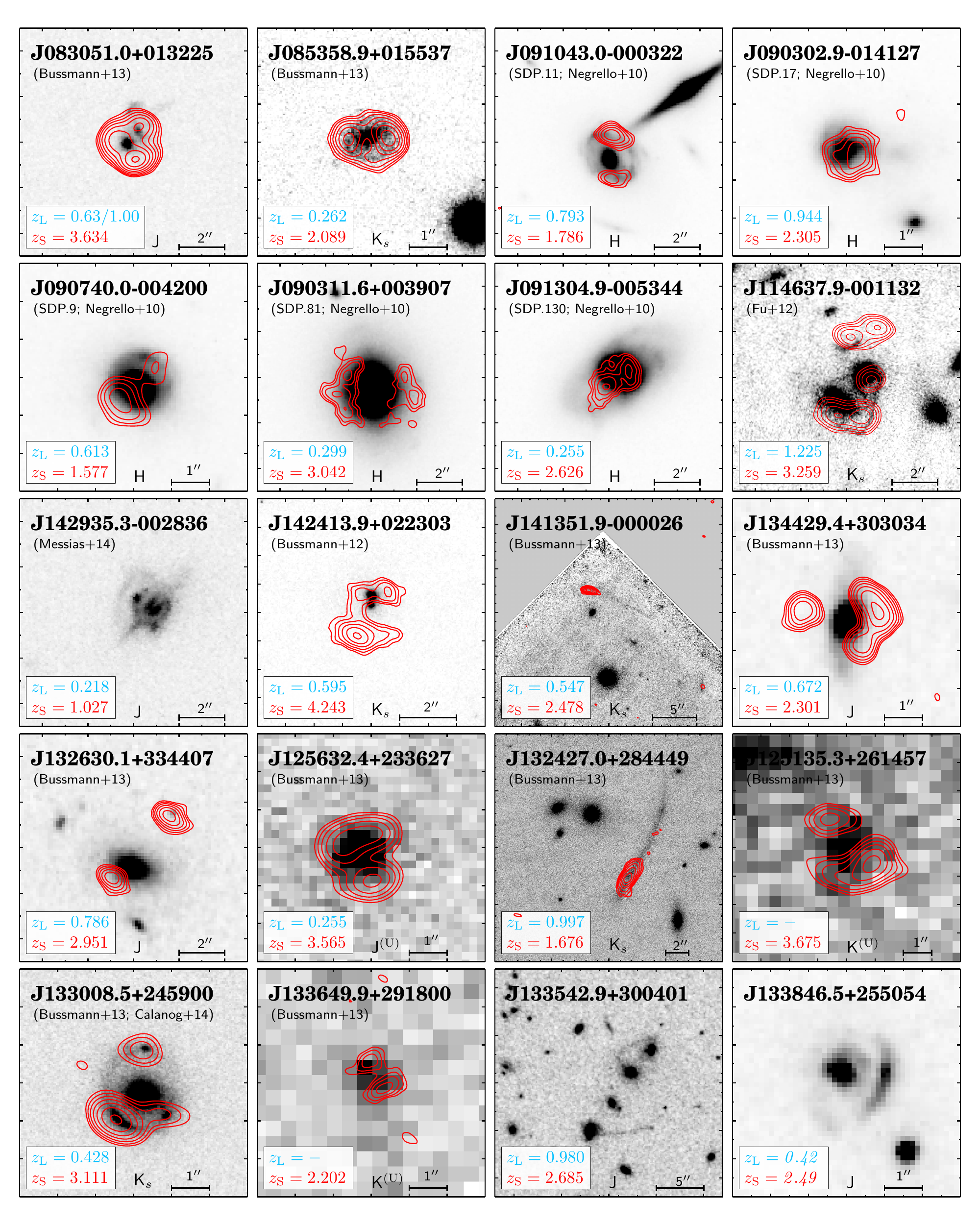}
 \vspace{-0.7cm}
 \caption{Postage stamps of
 the 20 confirmed lensed galaxies in the {\it H}-ATLAS fields. SMA
 880$\mu$m signal-to-noise ratio contours are superimposed to either the {\it HST}/WFC3/F110W (J-band), or {\it HST}/WFC3/F160W (H-band), 
or {\it Keck} K$_{s}$-band or UKIDSS J-/K-band image (the latter is denoted by the apex $^{\rm (U)}$). 
The redshifts of the lens, $z_{\rm L}$, and the redshift of the background galaxy, $z_{\rm S}$, are shown 
at the bottom of each stamp (in italic style when photometrically determined).}
 \label{fig:stamps_rankA}
\end{figure*}

The sample includes 20 sources that have already been confirmed to be strongly lensed 
via the detection of multiple images or arcs in the near infrared with {\it HST} and {\it Keck}/AO or, in the sub-millimeter, with the SMA,
and via the measurement of two distinct redshifts along the same line
of sight. 
They are shown in Fig.\,\ref{fig:stamps_rankA}. None of them lies in the SGP field because the {\it Herschel} data for that field were the latest to be delivered and processed, while 
most of the follow-up efforts have been focused on the other {\it H}-ATLAS fields. The confirmed lensed galaxies comprise $80\%$ of the sub-sample 
of candidate lensed galaxies with $F_{\rm 500}\ge 150\,$mJy identified in the equatorial and NGP fields (i.e. 17 out of 21). 
Among them are the first five lensed galaxies discovered by the {\it H}-ATLAS \citep{Neg10}.
The others are from \cite{Buss12,Buss13}, \cite{Fu12}, \cite{Messias14}, \cite{Calanog14}.
While most of them are galaxy-scale lenses, the sample includes also two group-lenses (HATLASJ114637.9-001132, HATLASJ133542.9+300401) 
as well as two cluster-lenses (HATLASJ141351.9-000026,
HATLASJ132427.0+284449) producing giant ($\sim10^{\prime\prime}$ in
near-IR images) arcs. 

The lenses in the sample exhibit a wide range of redshifts, from
$z_{\rm L}=0.22$ up to $z_{\rm L}=1.22$, with a median redshift
  $z_{\rm L}=0.6$. This is a consequence of our
selection technique, which 
is based exclusively on the observed properties of the background source, i.e. its flux density. No prior information on the lens properties is introduced during the lensed galaxy search. 
This means that samples of sub-mm selected lenses can allow us to study of the evolution of the mass density profile of galaxies out to high redshift ($z>1$) 
and over a wide range of galaxy masses \citep[e.g.][]{Dye14}. 

Magnification factors, $\mu$, have been derived from the lens modeling of the high resolution imaging data shown in Fig.\,\ref{fig:stamps_rankA}. Those obtained at sub-millimeter wavelengths 
from SMA (and ALMA for HATLAS090311.6+003907 and HATLASJ142935.3-002836) data 
are reported in Table\,\ref{tab:lens_candidates_F500ge100mJy}. For
three objects the magnification factor derived by \cite{Buss13} has
been replaced by a lower limit: 
it is the case of the two cluster-scale lenses, i.e. HATLASJ141351.9$−$000026 and HATLASJ132427.0+284449, which were modelled by Bussmann et al. as if they were galaxy-scale lenses, and SDP.17, for which SMA did not resolve the multiple images. 
The measured values\footnote{The magnification factors are derived
  by assuming either an analytic Sersic profile for the light profile of the
  background galaxy \citep{Buss13} or a pixelized surface brightness
  distribution \citep[e.g.][]{Dye14}.} of $\mu$ are in the range $\mu\sim5-15$, consistent with expectations \citep{Neg07}. As we will discuss in section\,\ref{sec:number_counts},  the statistics of lensed galaxies (i.e. the  sub-mm bright number counts) 
can also provide information on the typical magnification experienced by the background source population to be compared with the measured values of $\mu$. 

Only one object in our sample of candidate lensed galaxies,
HATLASJ084933.4+021442, has been so far confirmed to not be a strongly
lensed galaxy. It is indeed is a binary system of Hyper Luminous
Infrared Galaxies\footnote{One of the two HyLIRGs is claimed to be weakly lensed (i.e. $\mu\sim1.5$)
by a foreground lenticular galaxy.} (HyLIRGs) at $z=2.410$
reported by \cite{Ivison13}, which is believed to represent the early
stage in the formation of the core of a massive galaxy cluster.

\setcounter{table}{3}
\begin{landscape}
\begin{table}
  \vspace{-0.5cm}
  \begin{center}
    \small
\caption{List of candidate lensed galaxies with $F_{\rm 500}\ge100$\,mJy extracted from the $\sim$600\,deg$^{2}$ 
of the {\it H}-ATLAS fields. For each source the following information is  provided: {\it Herschel}/SPIRE flux densities; 
reliability of association with any SDSS source within 5$^{\prime\prime}$ from the SPIRE position and with  $r<22.4$ (Bourne et al. 2016; Furlanetto et al. 2016); distance of 
the SDSS source from the SPIRE detection and its AB $r$-band magnitude;
redshift of the SDSS association and of the SPIRE source (when no spectroscopic measurement is available, the photometric redshift is provided instead, in italic style); 
lensing rank: A={\it confirmed to be lensed}, B={\it likely to be lensed}, C={\it unclear}, D={\it not (strongly) lensed}; magnification factor derived from the modeling of high resolution sub-mm/mm imaging data.}
\label{tab:lens_candidates_F500ge100mJy}
    \vspace{-0.0cm}
    \tiny
\begin{tabular}{lrrrrrrrrrrr}   
\hline 
 \hline 
\multicolumn{1}{c}{H-ATLAS IAU name} & 
\multicolumn{1}{c}{$F_{\rm 250}$} & 
\multicolumn{1}{c}{$F_{\rm 350}$} &  
\multicolumn{1}{c}{$F_{\rm 500}$}  & 
\multicolumn{1}{c}{reliab.}   & 
\multicolumn{1}{c}{sep($^{\prime\prime}$)}   & 
\multicolumn{1}{c}{r$-$mag}   & 
\multicolumn{1}{c}{$z_{\rm opt}$}  & 
\multicolumn{1}{c}{$z_{\rm sub-mm}$}  &
\multicolumn{1}{c}{Lensing}   &
\multicolumn{1}{c}{$\mu_{\rm sub-mm}$} \\ 
                                                & 
\multicolumn{1}{c}{(mJy)}        & 
\multicolumn{1}{c}{(mJy)}         &  
\multicolumn{1}{c}{(mJy)}        &   
\multicolumn{1}{c}{SDSS-ID}  &  
\multicolumn{1}{c}{SDSS-ID}  &
\multicolumn{1}{c}{SDSS-ID}  &                                        
                                              &                                                                           
                                              & 
\multicolumn{1}{c}{rank}        &
    \\
\hline
\multicolumn{4}{l}{{\bf GAMA-9h field} (53.4\,deg$^{2}$)} \\
HATLASJ083051.0+013225{\tiny (B13)}             &  248.5$\pm$7.5    &  305.3$\pm$8.1  &  269.1$\pm$8.7   &  0.958   &   1.81   &   21.6    &  0.626+1.002               &  3.634$^{\ddagger}$      &  A   &  6.9$\pm$0.6 \\
HATLASJ085358.9+015537{\tiny (B13)}             &  396.4$\pm$7.6   &   367.9$\pm$8.2  &  228.2$\pm$8.9  &  -          &   -        &   -         & 0.262$^{\star}$              &  2.0925$^{\diamond}$  &  A  & 15.3$\pm$3.5 \\
HATLASJ091043.0$-$000322{\tiny (SDP.11)}    &  420.8$\pm$6.5   &    370.5$\pm$7.4  &  221.4$\pm$7.8  &  -          &   -        &   -         &  0.793                           &  1.786       &  A & 10.9$\pm$1.3 \\	
                                                                          &                               &                             &                             &   0.027   &  4.39    &  20.4    &  {\it 0.457$\pm$0.055}  &      &     &  \\
HATLASJ090302.9$-$014127{\tiny (SDP.17)}    &  354.1$\pm$7.2   &   338.8$\pm$8.1  &  220.0$\pm$8.6  &  0.988   &   0.83   &   21.7    &  0.9435            &  2.3049  &  A  & $>4.9$ \\	
HATLASJ084933.4+021442{\tiny (I13)}             &  216.7$\pm$7.5   &    248.5$\pm$8.2  &  208.6$\pm$8.6  &  0.777   &   1.65  &   20.4    &  -                                          &  2.410          &  D & \\	
HATLASJ090740.0$-$004200{\tiny (SDP.9)}      &  477.6$\pm$7.3   &    327.9$\pm$8.2  &  170.6$\pm$8.5  &  0.988   &   0.80   &   22.0    & 0.6129   &  1.577$^{e}$     &   A & 8.8$\pm$2.2 \\
HATLASJ090311.6+003907{\tiny (SDP.81)}       & 133.2$\pm$7.4    &   186.1$\pm$8.2  &  165.2$\pm$8.8   &  0.996   &   1.86   &   18.8    & 0.2999            &   3.042$^{e}$     &  A & 16.0$\pm$0.7 \\	
HATLASJ091840.8+023048{\tiny (B13)}            &  125.7$\pm$7.2   &   150.7$\pm$8.3  &   128.4$\pm$8.7  &  -          &   -        &   -        & -                                          &   2.581 &  C & \\	
HATLASJ091304.9$-$005344{\tiny (SDP.130)}  &  118.2$\pm$6.4    &   136.8$\pm$7.4  &   104.3$\pm$7.7  &   0.987  &  2.16    &  19.3     & 0.2201    &  2.6260   &  A & 2.1$\pm$0.3 \\
      \hline
\multicolumn{4}{l}{{\bf GAMA-12h field} (53.6\,deg$^{2}$)} \\
HATLASJ114637.9$-$001132{\tiny (F12)}   &   316.0$\pm$6.6   &   357.9$\pm$7.4    &   291.8$\pm$7.7    &  0.990     &  0.42     &  21.7    & 1.2247                   &   3.259                &  A & 7.6$\pm$1.5 \\   
                                                                    &                               &                                &                                &  0.000     & 4.61    &  21.3    &  {\it 0.470$\pm$0.095}  &      &     &  \\
HATLASJ113526.2$-$014606{\tiny (B13)}   &   278.8$\pm$7.4   &    282.9$\pm$8.2   &    204.0$\pm$8.6   &  -            &  -           & -         & -                                              &   3.1276  &  C & \\   
HATLASJ121334.9$-$020323                     &   211.0$\pm$6.5   &    197.9$\pm$7.5   &    129.9$\pm$7.7  &  0.999     &  0.20      & 18.7    & 0.190$^{\dagger}$ &  {\it 1.89$\pm$0.35}            &     B &  \\  
HATLASJ121301.5$-$004922                     &   136.6$\pm$6.6   &    142.6$\pm$7.4    &   110.9$\pm$7.7   &  0.972     &  1.51      & 21.6    & {\it 0.191$\pm$0.080}            &  {\it 2.35$\pm$0.40}            &   B &   \\  
HATLASJ120709.2$-$014702                     &   143.2$\pm$7.4   &    149.2$\pm$8.1   &    110.3$\pm$8.7  &  -            &  -           & -          & -                                              &   {\it 2.26$\pm$0.39}  &   C &   \\  
HATLASJ120319.1$-$011253                     &    114.3$\pm$7.4   &   142.8$\pm$8.2    &   110.2$\pm$8.6    &  -            &  -           & -          & -                                              &  {\it 2.70$\pm$0.44}  &   C &    \\  
HATLASJ115101.7$-$020024                     &    183.5$\pm$7.3   &    164.7$\pm$8.0   &    108.7$\pm$8.6   &  -            &  -           & -          & -                                              &  {\it 1.81$\pm$0.34}  &  C &    \\
HATLASJ115112.2$-$012637                     &    141.2$\pm$7.4   &    137.7$\pm$8.2   &    108.4$\pm$8.8  & 0.996     &  0.45      & 20.2     & 0.426$^{\dagger}$          &  {\it 2.22$\pm$0.39}  &  B &    \\
HATLASJ120127.6$-$014043                     &      67.4$\pm$6.5    &   112.1$\pm$7.4    &   103.9$\pm$7.7    &  -            &  -           & -         & -                                            &   {\it 3.80$\pm$0.58}  &  C &    \\
HATLASJ120127.8$-$021648                     &    207.9$\pm$7.3   &    160.9$\pm$8.2   &    103.6$\pm$8.7   &  -            &  -           & -          & -                                             &  {\it 1.50$\pm$0.30}  &  C  &    \\
HATLASJ121542.7$-$005220                     &    119.7$\pm$7.4   &    135.5$\pm$8.2   &    103.4$\pm$8.6   &  -            &  -           & -          &  -                                           &   {\it 2.48$\pm$0.42}  &  C  &   \\
HATLASJ115820.1$-$013752                     &    119.8$\pm$6.8   &    123.7$\pm$7.7   &    101.5$\pm$7.9   &  -            &  -           & -          & -                                             &  2.1911  &   C &   \\
\hline
\multicolumn{4}{l}{{\bf GAMA-15h field} (54.6\,deg$^{2}$)} \\
HATLASJ142935.3$-$002836{\tiny (M14)}    &  801.8$\pm$6.6  &  438.5$\pm$7.5   &   199.8$\pm$7.7   &  0.985       &  1.58        &  20.6     & 0.2184 &  1.027     &  A  &  10.8$\pm$0.7  \\   
HATLASJ142413.9+022303{\tiny (B12)}        &  112.2$\pm$7.3   &   182.2$\pm$8.2   &   193.3$\pm$8.5   &  0.986       &  0.74       &   22.1     &  0.595     &  4.243               &  A  &  4.6$\pm$0.5 \\    
HATLASJ141351.9$-$000026{\tiny (B13)}    & 188.6$\pm$7.4   &   217.0$\pm$8.1  &   176.4$\pm$8.7   &  0.980       &   1.12       &   22.2    &  0.5470  &  2.4782 &  A  &  $>1.8$ \\
HATLASJ144608.6+021927                          &  73.4$\pm$7.1     &   111.7$\pm$8.1  &   122.1$\pm$8.7   &  -              &  -             & -           &  -                                           &  {\it 4.10$\pm$0.61}  &   C &  \\	 	 
HATLASJ144556.1$-$004853{\tiny (B13)}    &  126.7$\pm$7.3   &   132.6$\pm$8.4  &   111.8$\pm$8.7   &  -              &  -             & -           &  -                                          &  {\it 2.51$\pm$0.42}  &  C &  \\
\hline
\multicolumn{4}{l}{{\bf NGP field} (170.1\,deg$^{2}$)} \\
HATLASJ134429.4+303034(B13)          &  462.0$\pm$7.4    &  465.7$\pm$8.6  &  343.3$\pm$8.7   &  0.987   &  0.43  &   21.9   &  0.6721                         &  2.3010      &  A   &  11.7$\pm$0.9 \\
HATLASJ132630.1+334407(B13)           & 190.6$\pm$7.3    &  281.4$\pm$8.2  &  278.5$\pm$9.0   &  0.922   &  2.16  &   21.3   & 0.7856                          &  2.951      &  A   &  4.1$\pm$0.3 \\
HATLASJ125632.4+233627(B13)          &  209.3$\pm$7.3    &  288.5$\pm$8.2  &  264.0$\pm$8.5   &  0.997   &  1.18  &   19.2   &  0.2551       &  3.565$^{\ddagger}$      &  A   &  11.3$\pm$1.7 \\
HATLASJ132427.0+284449(B13,G13)   &  342.4$\pm$7.3    &  371.0$\pm$8.2  &  250.9$\pm$8.5   &  -          &  -      &   -        &  0.997                            &  1.676      &  A &  $>2.8$ \\
HATLASJ132859.2+292326(B13)          &  268.4$\pm$6.5    &  296.3$\pm$7.4  &  248.9$\pm$7.7   &  -           &  -      &   -        &  -                                  &  2.778      &  C   &  \\
HATLASJ125135.3+261457(B13)          &  157.9$\pm$7.5    &  202.3$\pm$8.2  &  206.8$\pm$8.5   &  -           &  -      &   -        &  -                                  &  3.675      &  A   &  11.0$\pm$1.0  \\
HATLASJ133008.5+245900(B13,C14)   &  271.2$\pm$7.2    &  278.2$\pm$8.1  &  203.5$\pm$8.5   &  0.986   &  1.38  &  20.7   &  0.4276                          &  3.1112$^{\ddagger}$   &  A   &  13.0$\pm$1.5 \\
HATLASJ133649.9+291800(B13)          &  294.1$\pm$6.7    &  286.0$\pm$7.6  &  194.1$\pm$8.2   &  -           &  -       &   -       &  -                                   &  2.2024       &  A   &  4.4$\pm$0.8 \\
HATLASJ132504.3+311534                  &  240.7$\pm$7.2    &  226.7$\pm$8.2  &  164.9$\pm$8.8   &  0.749   &  2.42  &   22.3   &   {\it 0.58$\pm$0.11}    &  {\it 2.03$\pm$0.36}       &  B   &  - \\
HATLASJ125759.5+224558                  &  272.4$\pm$7.3    &  215.0$\pm$8.1  &  137.8$\pm$8.7   &  0.985   &  1.14  &   21.0   &  {\it 0.513$\pm$0.021}  &   {\it 1.53$\pm$0.30}       &  B   &  - \\
HATLASJ133846.5+255054                  &  159.0$\pm$7.4    &  183.1$\pm$8.2  &  137.6$\pm$9.0   &  0.965   &  1.95  &   20.7   & {\it 0.42$\pm$0.10}      &  {\it 2.49$\pm$0.42}       &  A   &  - \\
HATLASJ125652.4+275900                  &  133.9$\pm$7.5    &  164.1$\pm$8.2  &  131.8$\pm$8.9   &  0.925   &  1.20  &  21.6    &  -                                    &  {\it 2.75$\pm$0.45}       &  C   &  - \\
HATLASJ133413.8+260457                  &  136.1$\pm$7.2    &  161.1$\pm$7.8  &  126.5$\pm$8.4   &  -          &  -       &   -        &  -                                   & {\it 2.63$\pm$0.44}       &  C   &  - \\
HATLASJ133542.9+300401                  &  136.6$\pm$7.2    &  145.7$\pm$8.0  &  125.0$\pm$8.5   &  0.477   &  3.13  &  21.5    &  0.980$^{\ast}$             &  2.685$^{\ddagger}$       &  A   &  - \\
HATLASJ133255.6+265528                 &  192.5$\pm$7.4    &  167.4$\pm$8.1  &  116.6$\pm$8.6   &  0.034   &  4.32  &  19.6     & {\it 0.070$\pm$0.025}   & {\it 1.81$\pm$0.34}       &  C   &  - \\
HATLASJ132419.0+320752                 &    84.5$\pm$6.8    &  116.0$\pm$7.6  &  115.4$\pm$8.0   &  -          &  -       &   -         &  -                                    & {\it 3.54$\pm$0.54}       &  C   &  - \\
HATLASJ133255.7+342207                 &  164.3$\pm$7.5    &  186.8$\pm$8.1  &  114.9$\pm$8.7   &  -          &  -       &   -         &  -                                    &  {\it 2.17$\pm$0.38}      &  C   &  - \\
HATLASJ125125.8+254929                 &   57.4$\pm$7.4     &    96.8$\pm$8.2  &  109.4$\pm$8.8   &  0.970   &  1.62  &  21.2     &  {\it 0.62$\pm$0.10}      &  {\it 4.47$\pm$0.66}      &  B   &  - \\
HATLASJ134158.5+292833                 &   174.4$\pm$6.7   &  172.3$\pm$7.7  &  109.2$\pm$8.1   &  0.996   &  1.68  &   18.6   &  {\it 0.217$\pm$0.015}   &  {\it 1.95$\pm$0.35}      &  B   &  - \\
HATLASJ131540.6+262322                 &    94.1$\pm$7.4   &  116.1$\pm$8.2  &  108.6$\pm$8.7   &  -           &  -       &   -        &  -                                     & {\it 3.12$\pm$0.49}       &  C   &  - \\
HATLASJ130333.1+244643                &    99.0$\pm$7.2   &  111.5$\pm$8.2  &  104.5$\pm$8.7    &  -           &  -       &   -        &  -                                     & {\it 2.91$\pm$0.47}       &  C   &  - \\
HATLASJ133038.2+255128                &   175.8$\pm$7.4   &  160.3$\pm$8.3  &  104.2$\pm$8.8   &  0.983   &  0.792  &  21.6   &  {\it 0.20$\pm$0.15}       & {\it 1.82$\pm$0.34}      &  B   &  - \\
HATLASJ130118.0+253708                &   60.2$\pm$6.8   &  101.1$\pm$7.7  &  101.5$\pm$8.1    &  -            &  -        &   -        &  -                                    & {\it 4.08$\pm$0.61}       &  C   &  - \\
HATLASJ134422.6+231951                 &   109.6$\pm$7.9   &  98.3$\pm$9.1  &  101.4$\pm$9.2     &  -          &  -      &   -         &  -                                     & {\it 2.58$\pm$0.43}       &  C   &  - \\
\hline
\multicolumn{4}{l}{{\bf SGP field} (269.5\,deg$^{2}$)}  \\
HATLASJ012407.3$-$281434      &   257.5$\pm$6.0   &    271.1$\pm$6.0     &    203.9$\pm$6.8   &             &           &             &  -          &  {\it 2.31$\pm$0.40}      &  C   &  - \\     
HATLASJ013840.4$-$281855      &   116.3$\pm$5.7  &    177.0$\pm$6.0     &   179.4$\pm$7.1    &             &           &             &  -          &  {\it 3.86$\pm$0.58}      &  C   &  - \\         
HATLASJ232531.3$-$302235      &   175.5$\pm$4.3  &      227.0$\pm$4.7   &     175.7$\pm$5.7   &             &           &             &  -          &  {\it 2.80$\pm$0.46}      &  C   &  - \\      
HATLASJ232419.8$-$323926      &   213.0$\pm$4.4  &      244.2$\pm$4.8   &     169.4$\pm$5.8   &             &           &             &  -          &  {\it 2.37$\pm$0.40}      &  C   &  - \\         
HATLASJ010250.8$-$311723      &   267.9$\pm$5.2   &     253.1$\pm$5.7     &   168.1$\pm$7.1   &             &           &             &  -          &  {\it 1.92$\pm$0.35}      &  C   &  - \\        
    \end{tabular} 
  \end{center}
\end{table}
\end{landscape}

\setcounter{table}{3}
\begin{landscape}
\begin{table}
  \vspace{-0.5cm}
  \begin{center}
    \small
\caption{\it Continued.}
    \vspace{-0.0cm}
    \tiny
\begin{tabular}{lrrrrrrrrrrr}   
\hline 
 \hline                &                                                                           &                                                                           &   \multicolumn{1}{c}{rank}        \\
\multicolumn{1}{c}{H-ATLAS IAU name} & 
\multicolumn{1}{c}{$F_{\rm 250}$} & 
\multicolumn{1}{c}{$F_{\rm 350}$} &  
\multicolumn{1}{c}{$F_{\rm 500}$}  & 
\multicolumn{1}{c}{reliab.}   & 
\multicolumn{1}{c}{sep($^{\prime\prime}$)}   & 
\multicolumn{1}{c}{r$-$mag}   & 
\multicolumn{1}{c}{$z_{\rm opt}$}  & 
\multicolumn{1}{c}{$z_{\rm sub-mm}$}  &
\multicolumn{1}{c}{Lensing}   &
\multicolumn{1}{c}{$\mu_{\rm sub-mm}$} \\ 
                                                & 
\multicolumn{1}{c}{(mJy)}        & 
\multicolumn{1}{c}{(mJy)}         &  
\multicolumn{1}{c}{(mJy)}        &   
\multicolumn{1}{c}{SDSS-ID}  &  
\multicolumn{1}{c}{SDSS-ID}  &
\multicolumn{1}{c}{SDSS-ID}  &                                        
                                              &                                                                           
                                              & 
\multicolumn{1}{c}{rank}        & \\
\hline
HATLASJ000912.7$-$300807      &   352.8$\pm$5.4    &    272.6$\pm$6.1      &  156.1$\pm$6.8    &             &           &             &  -          &  {\it 1.39$\pm$0.29}     & C   &  - \\         
HATLASJ234418.1$-$303936      &   125.8$\pm$5.1  &      185.5$\pm$5.6   &     155.1$\pm$7.0   &             &           &             &  -          &  {\it 3.27$\pm$0.51}      &  C   &  - \\         
HATLASJ234357.7$-$351723      &   263.5$\pm$5.3   &     223.0$\pm$5.8    &    154.2$\pm$7.0   &             &           &             &  -          &  {\it 1.73$\pm$0.33}      &  C   &  - \\         	
HATLASJ002624.8$-$341737      &   137.7$\pm$5.2    &    185.9$\pm$5.8     &   148.8$\pm$6.8    &             &           &             &  -          & {\it  2.96$\pm$0.48}     &  C   &  - \\          	
HATLASJ012046.4$-$282403      &   103.3$\pm$5.7   &     149.8$\pm$5.8    &    145.7$\pm$7.4   &             &           &             &  -          &  {\it 3.62$\pm$0.55}      &  C   &  - \\          	
HATLASJ235827.6$-$323244      &   112.5$\pm$4.6   &     148.0$\pm$5.2   &     143.4$\pm$6.1    &             &           &             &  -          & {\it 3.37$\pm$0.52}     &  C   &  - \\          
HATLASJ225844.7$-$295124      &   175.4$\pm$5.2   &     187.0$\pm$5.9   &     142.6$\pm$7.5    &             &           &             &  -          &  {\it 2.36$\pm$0.40}      &  C   &  - \\      	
HATLASJ230815.5$-$343801      &   79.4$\pm$5.4   &     135.4$\pm$5.7     &   140.0$\pm$7.0   &             &           &             &  -          &  {\it 4.23$\pm$0.63}      &  C  &  - \\           
HATLASJ224805.3$-$335820      &   122.3$\pm$5.7   &     135.5$\pm$6.3     &   126.9$\pm$7.2   &             &           &             &  -          &  {\it 2.86$\pm$0.46}      &  C   &  - \\          
HATLASJ232623.0$-$342642      &   153.7$\pm$4.4    &    178.3$\pm$5.0     &   123.5$\pm$6.2   &             &           &             &  -          &  {\it 2.40$\pm$0.41}      &  C   &  - \\        
HATLASJ232900.6$-$321744      &   118.3$\pm$4.7    &    141.2$\pm$5.2    &    119.7$\pm$6.4   &             &           &             &  -          &  {\it 2.81$\pm$0.46}      &  C   &  - \\        
HATLASJ013239.9$-$330906      &   112.0$\pm$5.5    &    148.8$\pm$6.2    &    117.7$\pm$7.0     &             &           &             &  -          &  {\it 2.90$\pm$0.47}      &  C   &  - \\          
HATLASJ000007.4$-$334059      &   130.3$\pm$5.4    &    160.1$\pm$5.9     &   116.2$\pm$6.5     &             &           &             &  -          &  {\it 2.56$\pm$0.43}      &  C   &  - \\       
HATLASJ005132.8$-$301848      &   164.6$\pm$5.4   &     160.2$\pm$5.8     &   113.0$\pm$7.2     &             &           &             &  -          &  {\it 2.05$\pm$0.37}      & C   &  - \\    
HATLASJ225250.7$-$313657      &   127.4$\pm$4.2    &    138.7$\pm$4.9     &   111.4$\pm$5.9     &             &           &             &  -          &  {\it 2.49$\pm$0.42}      &  C   &  - \\      
HATLASJ230546.2$-$331038      &   76.8$\pm$5.6   &     110.9$\pm$5.9     &   110.4$\pm$7.0     &             &           &             &  -          &  {\it 3.67$\pm$0.56}      &  C   &  - \\       
HATLASJ000722.1$-$352014      &   237.3$\pm$5.2      &  192.8$\pm$5.6       & 107.5$\pm$6.6     &             &           &             &  -         &  {\it 1.46$\pm$0.30}      &  C   &  - \\     	
HATLASJ013951.9$-$321446      &   109.0$\pm$4.9    &    116.5$\pm$5.3      &  107.1$\pm$6.2    &             &           &             &  -         &  {\it 2.73$\pm$0.45}      &  C   &  - \\         
HATLASJ003207.7$-$303724      &   80.3$\pm$5.0      &  106.2$\pm$5.2      &  105.8$\pm$6.3      &             &           &             &  -          &  {\it 3.45$\pm$0.53}      &  C   &  - \\	
HATLASJ004853.2$-$303109      &   118.1$\pm$4.5    &    147.3$\pm$5.0      &  105.4$\pm$6.0    &             &           &             &  -          &  {\it 2.59$\pm$0.43}      &  C  &  - \\         
HATLASJ005132.0$-$302011      &   119.3$\pm$5.0    &    121.0$\pm$5.8    &    102.0$\pm$6.6    &             &           &             &  -         &  {\it 2.42$\pm$0.41}      &  C   &  - \\       
HATLASJ224207.2$-$324159      &   73.0$\pm$5.5   &      88.1$\pm$6.2    &    100.8$\pm$7.7     &             &           &             &  -          &  {\it 3.60$\pm$0.55}     &  C  &  - \\         
HATLASJ013004.0$-$305513      &   164.4$\pm$4.3      &  147.5$\pm$5.1     &   100.6$\pm$5.9     &             &           &             &  -          &  {\it 1.84$\pm$0.34}      &  C   &  - \\       	
HATLASJ223753.8$-$305828      &   139.1$\pm$4.9   &     144.9$\pm$5.1   &     100.6$\pm$6.2    &             &           &             &  -          &  {\it 2.17$\pm$0.38}      &  C   &  - \\          
HATLASJ012415.9$-$310500      &   140.4$\pm$5.4     &   154.5$\pm$5.7     &   100.3$\pm$7.0     &             &           &             &  -          &  {\it 2.20$\pm$0.38}      &  C   &  - \\   
\hline
\hline
\multicolumn{11}{l}{B13 = Bussmann et al. (2013); I13 = Ivison et
  al. (2013); F12 = Fu et al. (2012); B12 = Bussmann et al. (2012);
  M14 = Messias et al. (2014);} \\
\multicolumn{11}{l}{ G13 = George et al. (2013); C14 = Calanog et al. (2014).} \\
\multicolumn{2}{l}{$^{\diamond}${\scriptsize from PdBI (Yang et al. 2016)}} \\
\multicolumn{2}{l}{$^{\star}${\scriptsize from SALT (Marchetti et al. in prep.)}} \\
\multicolumn{2}{l}{$^{\dagger}${\scriptsize from NTT (Amber 2015)}} \\
\multicolumn{2}{l}{$^{\ddagger}${\scriptsize from CARMA (Riechers et al. in prep.)}} \\
\multicolumn{2}{l}{$^{\ast}${\scriptsize Stanford et al. 2014, ApJ, 213, 25}} \\
    \end{tabular} 
  \end{center}
\end{table}
\end{landscape}

%

\subsection{Colours and redshifts}\label{subsec:colours_redshifts}

A colour-colour plot based on the {\it Herschel}/SPIRE photometry is
shown in Fig.\,\ref{fig:colorcolor_plot} for candidate lensed galaxies (dots),
local galaxies (squares) and blazars (triagles) with $F_{\rm 500}\ge100\,$mJy. In the same
figure we have delimited the regions occupied by objects with spectral energy distribution (SED)
peaking at $\sim350\,\mu$m (i.e. with $F_{\rm 250}$/$F_{\rm 350}<1$
and $F_{\rm 350}$/$F_{\rm 500}>1$; ``350\,$\mu$m peakers'') and by
objects with SED rising from 250 to 500$\,\mu$m (i.e. with $F_{\rm
  250}\le$F$_{\rm 350}\le F_{\rm 500}$; ``500$\,\mu$m risers''). There
is clearly a bimodal distribution in the $F_{\rm 250}/F_{\rm 350}$ values, with
candidate lensed galaxies having significantly ``redder'' colours (i.e. higher emission at 350$\,\mu$m than at 250$\,\mu$m) compared
to local galaxies. This is consistent with the former
lying at much higher redshift (i.e. $z\gsim1$). In fact, the majority
of the candidate lensed galaxies are classified as 350$\,\mu$m peakers. 
The candidate lensed galaxy with the ``bluest'' colors, i.e. with $F_{\rm 250}/F_{\rm 350}>1.7$, is HATLASJ142935.3-002836, 
our lowest redshift ($z=1.027$) confirmed lensed source \citep[][see fig.\,\ref{fig:stamps_rankA}]{Messias14}.

\cite{Pearson13} have used a sub-set 
of 40 {\it H}-ATLAS sources with spectroscopic redshifts in the range $0.5<z<4.2$ to construct an empirical SED 
for high-redshift {\it Herschel} selected sources to be used to
estimate photometric redshifts from {\it Herschel}/SPIRE data. 
The template is the sum of two modified black body spectra with temperatures 
$T_{\rm cold}=23.9\,$K and $T_{\rm hot}=46.9\,$K, and dust emissivity index fixed to $\beta=2$. 
The ratio between the normalization of the two components is 30.1. 
In fig.\,\ref{fig:colorcolor_plot} we show the track of the empirical SED for redshifts from 0.5 to 4.5 in steps of 0.5. 
The  colours of the candidate lensed galaxies are consistent with them being in the redshift range $1.5-4.5$.
Their photometric redshifts, derived by fitting the {\it Herschel}/SPIRE photometry with the Pearson et al. template, 
are listed in Table\,\ref{tab:lens_candidates_F500ge100mJy}, in italic style. 
The errors are calculated as $0.12\times(1+z_{\rm phot})$,  0.12 being
the rms scatter in the $(z_{\rm phot} - z_{\rm spec})/(1+z_{\rm spec})$ values as measured by
Pearson et al. for sources with $z_{\rm spec}>1$.

\begin{figure}
  \hspace{-0.0cm}\includegraphics[width=8.3cm]{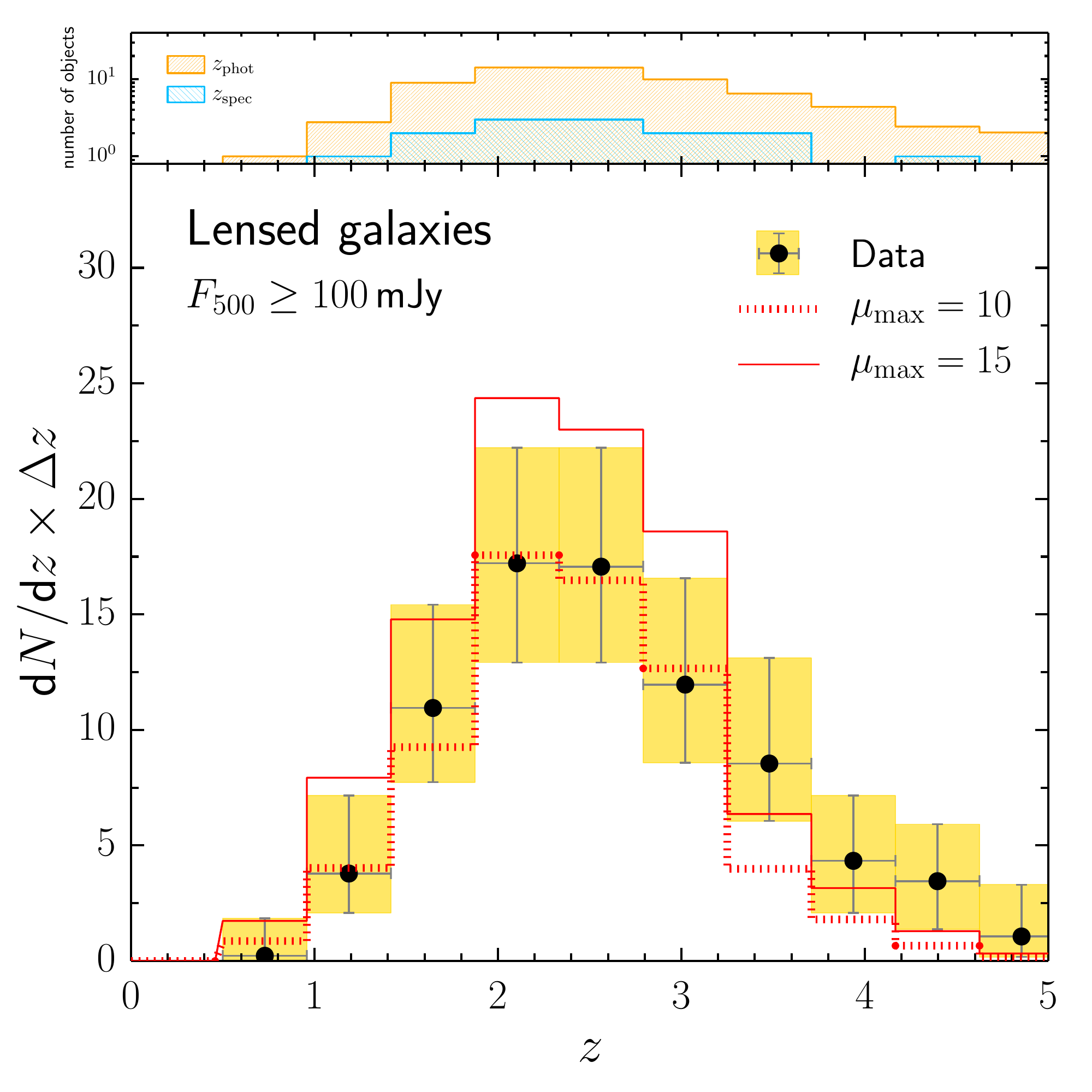}
 \vspace{-0.3cm}
 \caption{Redshift distribution of the candidate lensed galaxies with F$_{\rm
     500}\ge100$\,mJy identified in the {\it H}-ATLAS. {\it Upper
     panel}:  distribution of the photometric redshift sample (orange)
   and of the
   spectroscopic redshift sample (blue).  {\it Lower panel}: distribution of
   the photometric+spectroscopic redshift sample (dots with error bars). The shaded region represents the 68\% confidence interval
   assuming Poisson statistics (Gehrels 1986). The
   curves are predictions based on the galaxy formation and evolution model of Cai et al. (2013) 
coupled with the lensing formalism of Lapi et al. (2012), assuming different maximum magnifications: $\mu_{\rm max}=10, 15$. 
}
 \label{fig:Nz_lensed_galaxies}
\end{figure}

In the lower panel of Fig.\,\ref{fig:Nz_lensed_galaxies} we show the redshift distribution of
the candidate lensed galaxies with $F_{\rm 500}\ge100\,$mJy, with
photometric redshifts replaced by spectroscopic ones where
available. The \cite{Ivison13} source is  not included in the plot as
it is known to not be strongly lensed. 
The distribution is derived by performing 10000 simulations.
Each time the redshift of the sources is resampled at random from a
Gaussian probability distribution with a mean equal to the measured
photometric/spectroscopic redshift and dispersion  equal to the
associated error. The simulated redshifts are then binned into a
histogram and the mean value in each bin is taken as the estimate of
the number of objects in that bin. Error bars, corresponding to the
68\% confidence interval, are derived following the prescriptions of \cite{G86} for Poisson statistics.
The median redshift is $z_{\rm median}=2.53$. 
Similar median redshifts are obtained for the spectroscopic
  redshift sample and the photometric redshift sample alone,
  i.e. $z_{\rm median}=2.49$ and $z_{\rm median}=2.54$, respectively
  (the two distributions are shown in the upper panel of fig.\,\ref{fig:Nz_lensed_galaxies}). 
The measured redshift distribution is consistent with predictions based
on the physical model of \cite{Cai13} once combined with the lensing formalism of \cite{Lapi12} for maximum magnifications in 
the range 10-15 (see section\,\ref{subsec:modelling} for details on the modeling). 

\begin{figure}
  \hspace{0.0cm}
  \begin{minipage}[b]{0.51\linewidth}
    \centering \resizebox{2.0\hsize}{!}{
      \hspace{-1.0cm}\includegraphics{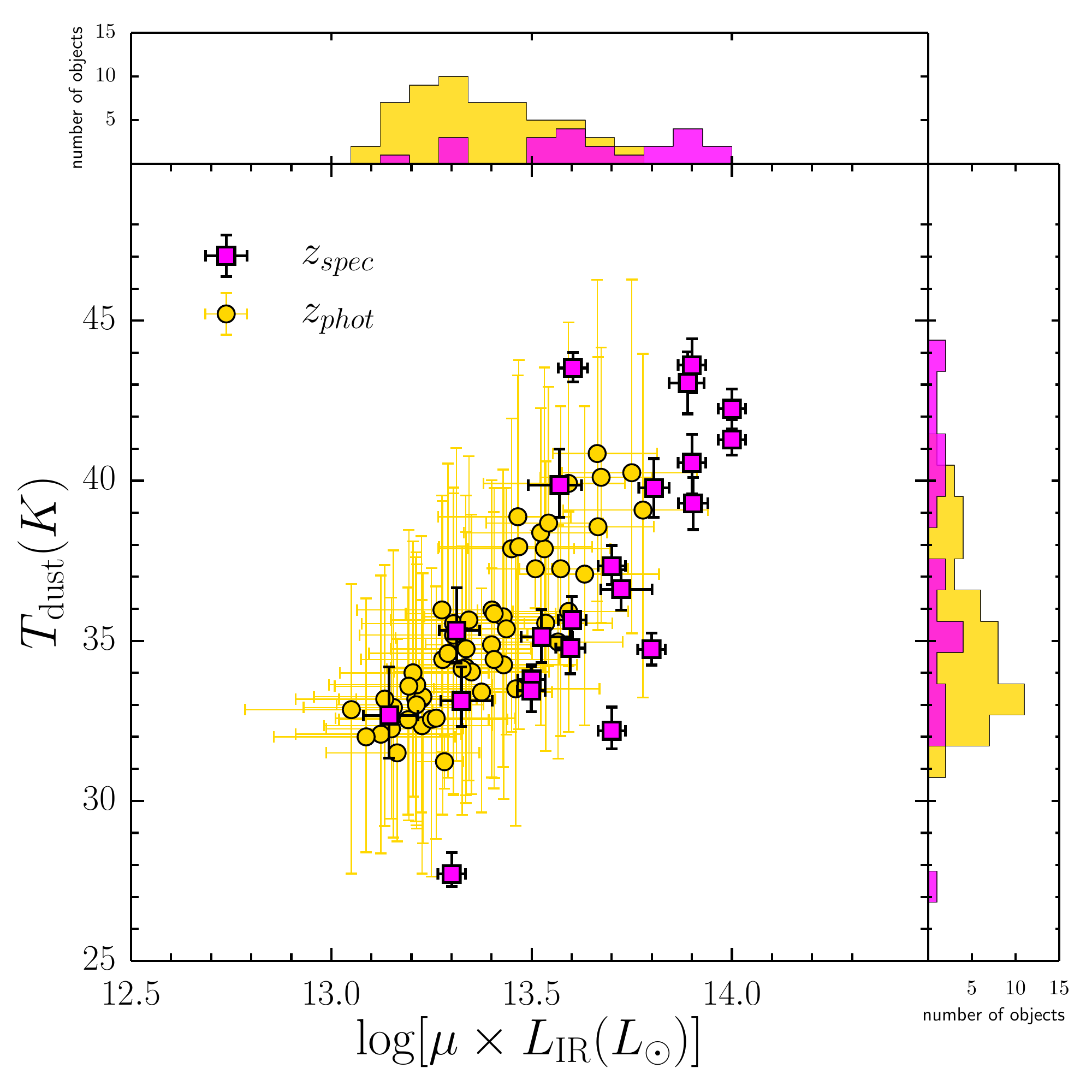}} 
  \end{minipage}
 \vspace{-0.3cm}
 \caption{Dust temperature versus infrared luminosity ($8-1000\,\mu$m, not corrected for lensing) of the candidate lensed galaxies extracted from the {\it H}-ATLAS fields, 
excluding the Ivison et al. (2013) source. Squares correspond to objects with spectroscopic redshift while
the dots indicate objects with photometric redshift. Error bars correspond to the 68\% confidence interval and take into account 
uncertainties on both the redshifts and the {\it Herschel}
photometry. The infrared luminosity and the dust temperature are
derived assuming an optically-thin greybody model with dust emissivity index fixed to $\beta=1.5$.} 
 \label{fig:Tdust_vs_lgLir_lensed_galaxies}
\end{figure}

We use the redshift information to derive the infrared luminosity, $L_{\rm IR}$ (integrated over the rest-frame wavelength range $8-1000\,\mu$m)
of our candidate lensed galaxies. We assume a single temperature,
  optically-thin, modified black body spectrum with dust emissivity index $\beta=1.5$
\citep[][]{Buss13,Nayyeri16} and we fit it to the {\it Herschel} photometry of each source.
The infrared luminosity and dust temperature, $T_{\rm dust}$, are kept as free parameters. In order to account for 
the uncertainty on the photometric/spectroscopic redshift, as well as that on the photometry, we perform for each source 1000 simulations by resampling at random 
the distributions of redshift values and measured flux densities, both assumed to be Gaussian. The final values of $L_{\rm IR}$ and $T_{\rm dust}$, shown 
in Fig.\,\ref{fig:Tdust_vs_lgLir_lensed_galaxies}, are obtained as the median of the derived distributions of 1000 best fit values. All the galaxies in the sample appear to be hyper luminous infrared
galaxies (HyLIRGs; i.e. $L_{\rm IR}\ge10^{13}\,L_{\odot}$). However, with an
expected typical magnification of $\sim5-15$ \citep[][see also Fig.\,\ref{fig:mu_vs_F500_mumax15}]{Lapi12,Buss13,Dye14} most of
them are likely to be {\it intrinsically} ultra luminous infrared galaxies (ULIRGs;
$10^{12}\le L_{\rm IR}/L_{\odot}\le 10^{13}$). 
The median value of the dust temperature is $T_{\rm
  dust}=34.6$, consistent with what found for 
other sub-mm/mm selected unlensed \citep{Magnelli12} and lensed
\citep{Weib13,Canameras15,Nayyeri16} galaxies. 

%
%
%

\subsection{Optical and near-infrared counterparts}

\cite{Bourne16} and \cite{Furlanetto16} used a likelihood-ratio (LR)
technique \citep{Richter1975,SS1992,Ciliegi2003} to identify SDSS counterparts with $r<22.4$ to the {\it
  H}-ATLAS sources in the GAMA fields and in the NGP field, respectively. 
This technique exploits the knowledge of the intrinsic positional uncertainty of the sources as well as the magnitude 
distributions of true counterparts and background sources to assign a
probability (``reliability'', $R$) to each potential match, thus
allowing to distinguish 
robust counterparts from chance allignments with background sources.
We consider an optical source to be reliable counterpart to the {\it Heschel}/SPIRE detection if $R>0.8$. 
We do not expect lensed DSFGs to be visible in the
optical above the adopted $r$-band limit, because of the combined
effect of high redshift and dust obscuration. However, in gravitational
lensing systems, the foreground object acting as the lens, usually an
elliptical galaxy at redshifts $z\sim0.3-0.8$ (see Table\,\ref{tab:lens_candidates_F500ge100mJy}), is likely to be detected at
optical and/or near-IR wavelengths. We find that 9 of the yet to be
confirmed lensed galaxies have a reliable optical counterpart in SDSS,
but the photometric redshift of the SDSS source (where available) is much lower than the
one derived from the {\it Herschel}/SPIRE photometry, as reported in Table\,\ref{tab:lens_candidates_F500ge100mJy}. This is
consistent with the lensing scenario where the lens is detected in the
optical while the background galaxy entirely contributes to the far-IR/sub-mm
emission. The estimated high reliability of the association between
two distinct objects is actually determined by gravitational lensing,
which physically relates them. This effect was already observed in the
first sample of 5 lensed galaxies discovered by \cite{Neg10}.

In Appendix\,\ref{app:stamps} we show optical (SDSS or KiDS) to near-IR
(UKIDSS-LAS or VIKING or HST/F110W where available) postage
stamps images of the yet to be confirmed lensed galaxies. The spatial resolution ranges from 0.14$^{\prime\prime}$ for HST/F110W, to $\lsim0.7^{\prime\prime}$ for KiDS $r$-band, 
1.4$^{\prime\prime}$ for SDSS $r$-band, and$\lsim1^{\prime\prime}$ for VIKING and UKIDSS-LAS.
The reliable
optical counterparts are marked in the optical $r$-band
images. No likelihood-ratio analysis has been performed
on near-IR catalogues yet, however we note that most of our candidate
lensed galaxies have a near-IR source (or even more than one in some
cases) within a 5$^{\prime\prime}$ distance, likely to be the lens, with eventually some contribution from the background galaxy
\citep[see e.g.][]{Neg14}. There is an ongoing effort
to measure the redshift of these near-IR counterparts to confirm that
they lie at $z\lsim1$.

Based on available imaging and spectroscopic
information we provide in Table\,\ref{tab:lens_candidates_F500ge100mJy}  a {\it lensing rank} for each
object as follows:
\begin{itemize}
\item[{\bf A}] = {\it The Herschel/{\rm SPIRE} source is confirmed to be strongly lensed} according to the detection
      of multiple images or arcs with {\it HST}/{\it Keck}/SMA and/or the availability of spectroscopic redshifts for both the
      lens and the background galaxy indicating the presence of two
      distinct objects along the line of sight;
\item[{\bf B}] = {\it The Herschel/{\rm SPIRE}  source is likely to be
    lensed} based on the difference in redshift between the optical/near-IR ID and the sub-mm detection
      although at least one of the two redshifts is photometric;
\item[{\bf C}] = {\it The nature of the Herschel/{\rm SPIRE}  source
    is unclear} because of the lack of a reliable optical/near-IR counterpart which may indicate that the
  lens is particularly faint and/or lying at $z>1$, or that the sub-mm
  source is an un-lensed HyLIRG or a cluster of HyLIRGs;
\item[{\bf D}] = {\it The Herschel/SPIRE source is not strongly lensed}.
\end{itemize}
To summarize, in our sample there are 20 confirmed lensed galaxies
(i.e. rank A), 8 sources have been assigned rank B, while 51 have rank
C. Only one object has been proved, so far, to not be strongly lensed (i.e. rank D).

\section{Number counts}\label{sec:number_counts}

In this Section we derive the number counts of the candidate lensed galaxies
and compare them with model predictions. 

\begin{table}
  \begin{center}
    \caption{Integral number counts of candidate lensed galaxies
      with $F_{500}\ge100$\,mJy in the {\it H}-ATLAS
      fields. The quoted errors on the number counts correspond to the
      68\% confidence interval.}\label{tab:number_counts_lensed}
    \vspace{-0.0cm}
    \begin{tabular}{cc}
      \hline 
      \hline 
      $\log{[F_{\rm 500}/{\rm mJy}]}$ & $N(>F_{\rm 500})$ \\
   & $\times$10$^{-3}$ (deg$^{-2}$) \\
\hline
2.000  & 131$_{-15}^{+17}$  \\
2.038  & 96$_{-13}^{+14}$  \\
2.075  & 73$_{-11}^{+13}$  \\
2.113  & 62$_{-10}^{+12}$  \\
2.150  & 53$_{-9}^{+11}$ \\
2.188  & 47$_{-9}^{+11}$ \\
2.225  & 38.3$_{-7.9}^{+9.8}$ \\
2.263  & 28.3$_{-6.8}^{+8.7}$ \\
2.300  & 25.0$_{-6.4}^{+8.3}$ \\
2.338  & 16.6$_{-5.2}^{+7.1}$ \\
2.375  & 11.6$_{-4.3}^{+6.3}$ \\
2.413  & 8.3$_{-3.6}^{+5.6}$ \\
2.450  & 3.3$_{-2.1}^{+4.4}$ \\
2.488  & 1.7$_{-1.4}^{+3.8}$ \\
2.525  & 1.7$_{-1.4}^{+3.8}$ \\
\hline
\hline
    \end{tabular} 
  \end{center}
\end{table}

\begin{figure*}
 \vspace{-3.0cm}
  \hspace{-0.0cm}\includegraphics[width=16cm]{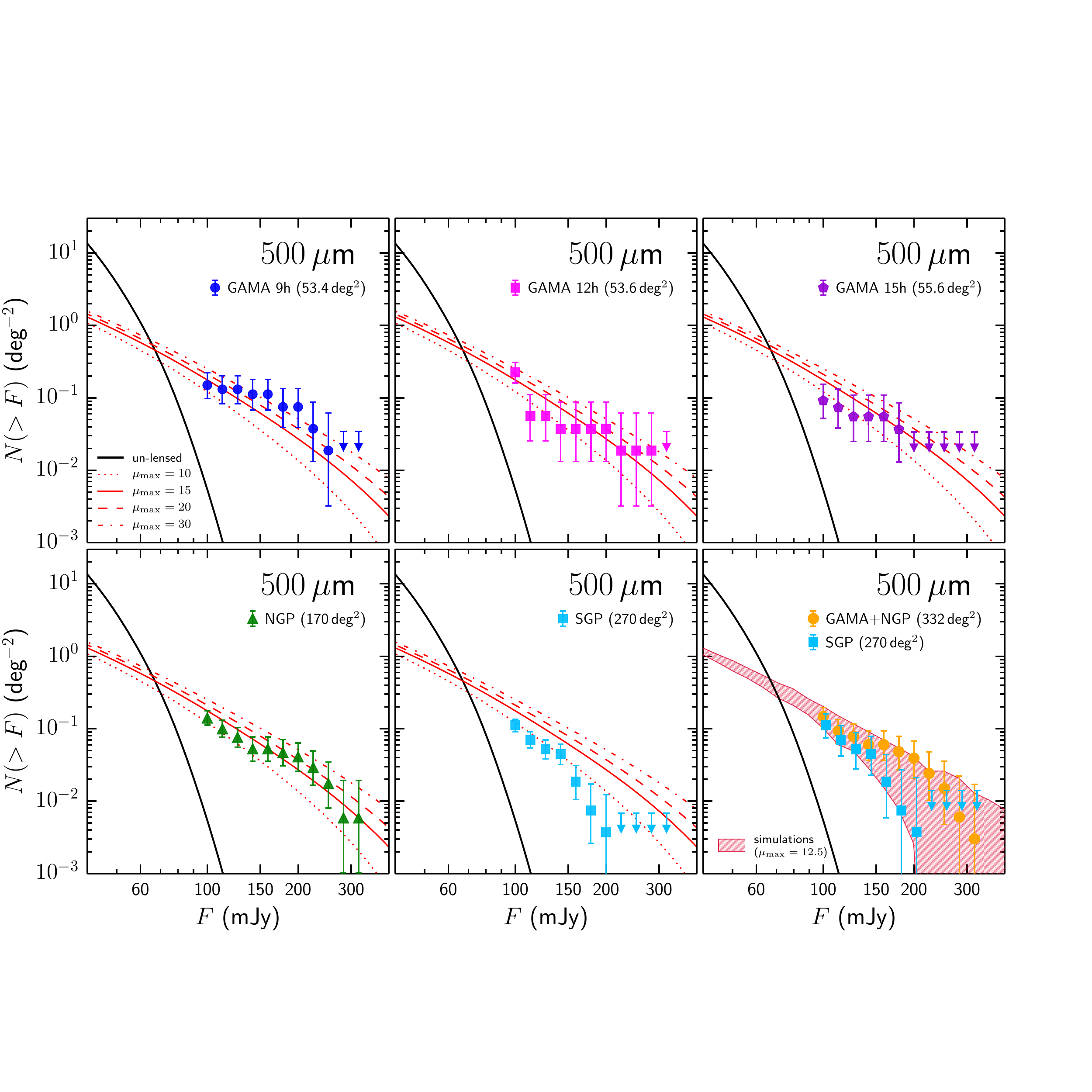}
 \vspace{-2.3cm}
 \caption{Integral number counts
   of the candidate lensed galaxies with $F_{\rm
     500}\ge100\,$mJy identified in the 5 {\it
     H}-ATLAS fields: GAMA\,9h (top-left), GAMA\,12h
   (top-middle), GAMA\,15h (top-right), NGP (bottom-left), and SGP
   (bottom-middle). The bottom-right panel shows the comparison between
   the number counts measured in the SGP field and those derived by
   combining the 3 GAMA fields with the NGP fields. Error bars and upper
   limits correspond to
   the 68\% confidence interval assuming Poisson statistic (Gehrels
   1986), with the exception of the bottom-right panel where they are
   shown at the 95\% confidence level. 
In all panels the solid black curve is the prediction for the abundance of un-lensed DSFGs at $z\gsim1.5$, based on the model of Cai
   et al. (2013). The red curves show the predicted number counts of
   lensed galaxies for different values of the maximum
   magnification, $\mu_{\rm max}$, experienced by the background sources, from 10 to
   30. The mass profile of the lenses is assumed to be the
   superposition of a NFW profile (for the dark matter component) and
   a de Vaucouleurs  profile (accounting for baryons), as described in Lapi et
   al. (2012). More details are given in
   section\,\ref{subsec:modelling}. The shaded region in the
   bottom-right panel illustrates the variation in the number density of
   lensed galaxies based on a set of 100 simulations performed over an area of
   300\,deg$^{2}$ and assuming $\mu_{\rm max}=12.5$.}
 \label{fig:counts_lensed_galaxies_HATLAS}
\end{figure*}

\subsection{Measurements}
 
The integral number counts of the candidate lensed galaxies  are
  shown in 
  Fig.\,\ref{fig:counts_lensed_galaxies_HATLAS} for each of the {\it
    H}-ATLAS fields: top panels for the
  three equatorial GAMA fields; bottom-left and bottom-middle panels for
  the NGP and SGP fields, respectively.  Error bars and upper limits correspond to the
68\% confidence interval, assuming Poisson statistics. The GAMA\,9h field and the
NGP field appear to be particularly rich in lensed galaxies with $F_{\rm
  500}\gsim200$\,mJy. On the contrary, the SGP field is
short of such bright sources. In fact, only one candidate
lensed galaxy with $F_{\rm 500}\geq200\,$mJy is found in the SGP field, while
13 have been identified in the other fields (most of which are already
confirmed lensed galaxies). 
In the bottom-right panel of the same figure we show the comparison
between the number counts extracted from the SGP field and those
derived by combining the 3 equatorial fields with the NGP
field. In this case the error bars and the upper limits have been
drawn to represent the 95\% confidence interval. It is clear that, despite the large difference in number density
for $F_{\rm 500}\gsim150\,$mJy, the two samples are
still consistent with each other at the $\sim$2\,$\sigma$ level. The observed
fluctuations are also consistent with the results of simulations based
on the modelled lensed number counts, as discussed in the
section\,\ref{subsec:modelling}. 

Once we combine the samples from all the 5 {\it H}-ATLAS different fields, the measured number
counts, shown in 
Fig.\,\ref{fig:counts_lensed_galaxies_HATLAS_vs_HeLMSplusHerS} and
reported in Table\,\ref{tab:number_counts_lensed}, are in very good
agreement 
with those derived from HeLMS+HerS \citep{Nayyeri16} for $F_{\rm
  500}\gsim125\,$mJy. At fainter flux densities, 
Nayyeri et al. report 77 candidate lensed galaxies with $F_{\rm 500}\geq100\,$mJy over an area of
370\,deg$^{2}$, to be compared with the 79 we find over
the 600\,deg$^{2}$ of the full {\it H}-ATLAS. 
This difference in number density, i.e. 0.21\,deg$^{-2}$ vs
0.13\,deg$^{-2}$, is still consistent with Poisson fluctuations at the
$\sim2\,\sigma$ level.

\subsection{Modelling}\label{subsec:modelling}

Given the {\it true} number density of un-lensed DSFGs per unit
logarithmic interval in flux density, $(dN/d{\rm lg}F)_{\rm T}$, the corresponding number density of lensed DSFGs is 
computed as \citep[e.g.][]{Lima2010,Lapi12}
\begin{eqnarray}
\left(\frac{dN}{d{\rm lg}F}\right)_{\rm L}(F) = \int_{\mu_{\rm min}}^{\mu_{\rm max}} \left( \frac{dN}{d{\rm lg}F} \right)_{\rm T} (F/\mu) p(\mu) d\mu,
\label{eq:lensed_counts1}
\end{eqnarray}
where $F$ is the {\it measured} flux density of the sources, and
$p(\mu)$ is the probability that a source has its flux boosted by a
factor $\mu$ as an effect of lensing. 

In general, the probability distribution of the magnifications depends
on the redshift of the source, $z_{\rm S}$, with higher redshift
objects being more likely lensed than lower redshift ones (as a
consequence of the higher optical depth for lensing). Therefore, a more accurate version of Eq.\,(\ref{eq:lensed_counts1}) is
\begin{eqnarray}
\left( \frac{dN}{d{\rm lg}F dz_{\rm S}} \right)_{\rm L} (F, z_{\rm S}) = ~~~~~~~~~~~~~~~~~~~~~~~~~~~~~~~~~~~~~~~ \nonumber \\
~~~~~~~~~~~~~~~ \int_{\mu_{\rm min}}^{\mu_{\rm max}} \left(
  \frac{dN}{d{\rm lg}F dz_{\rm S}} \right)_{\rm T} (F/\mu) p(\mu, z_{\rm S}) d\mu,
\label{eq:ciccio}
\end{eqnarray}
with the number densities now given per unit interval in redshift (of
the source) as well as per unit logarithmic interval in flux
density. Here we focus on the {\it strong} lensing regime, which
occurs when multiple images of the background source are formed. The
typical separation between multiple images is of the order of one to a few
arcseconds. Such angular scales are below the resolution capabilities
of {\it Herschel} which, in case of a lensing event, can only measure
the summed flux of the multiple images. For this reason we will only
consider the {\it total} magnification, $\mu_{\rm tot}$,
experienced by a source, i.e. the sum of the modulus of the
magnifications of the individual images. In Eq.\,(\ref{eq:ciccio}), the minimum value of $\mu$ in the
integration is fixed to the lowest value of $\mu_{\rm tot}$ that
corresponds to the formation of multiple images. For lenses with a
mass profile described by a
Singular Isothermal Sphere (SIS) model $-$ as supported by the
modelling of individual lensing systems \citep[e.g.][]{Dye14} $-$ the strong lensing regime is
achieved when $\mu_{\rm tot}\geq2$. Therefore, we set $\mu_{\rm
  min}=2$. On the other hand, the value of $\mu_{\rm max}$ mainly depends on the
intrinsic angular size of the background source: in general, the more compact the
source the higher the maximum magnification that can be achieved \citep[see e.g.
fig.\,8 of][]{Lapi12}. \cite{P02}
showed that for a source in the redshift range $z=1-4$, with typical
size $\simeq$1-10\,kpc, consistent with observations of the dust
emitting region in DSFGs \citep[e.g.][]{Riechers13,Riechers14,DeBreuck14,Simpson14,Simpson15,Ikarashi15,Hodge15,Hodge16,Lindroos2016}, the value of $\mu_{\rm tot}$ lies in the range
10 to 30 \citep[see also][]{Lapi12}. 
In the following, we work out predictions for different values for $\mu_{\rm
  max}$ in the range 10 to 30.

\begin{figure}
  \hspace{-0.4cm}\includegraphics[width=8.7cm]{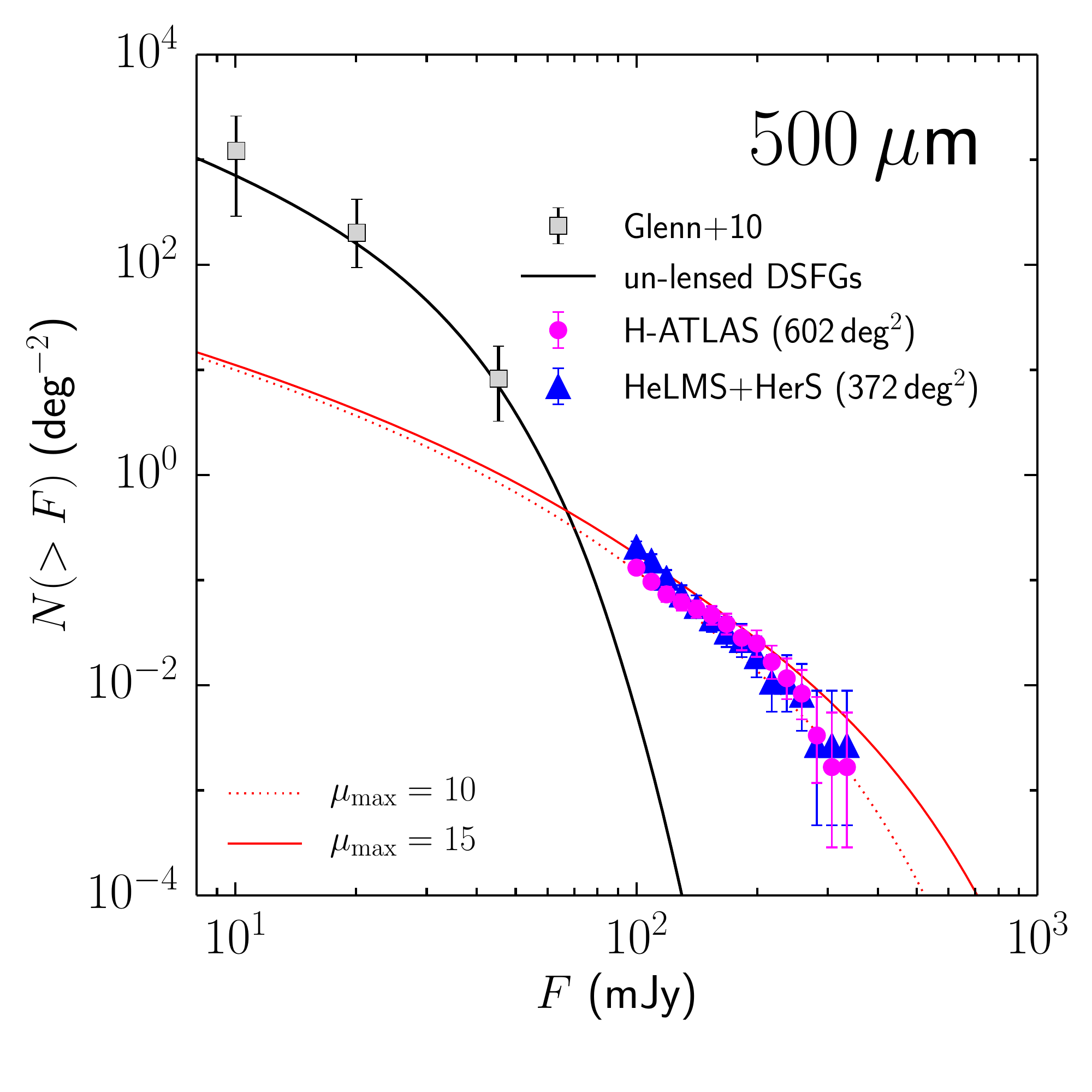}
 \vspace{-0.3cm}
 \caption{Integral number counts
   of the candidate lensed galaxies with $F_{\rm
     500}\ge100\,$mJy identified in all the {\it
     H}-ATLAS fields (purple dots) compared with the number counts of candidate
   lensed galaxies derived by Nayyeri et al. (2016) from HeLMS+HerS
   (blue triangles). Error bars correspond to the 95\,\% confidence
   interval. For clarity, a small offset in flux density have
   been applied to the {\it H}-ATLAS data. The meaning of the curves is the same as in
   fig.\,\ref{fig:counts_lensed_galaxies_HATLAS}. The squares represent
   measurements of the number density of un-lensed DSFGs derived from P(D)
   analysis \citep{Glenn10}.}
 \label{fig:counts_lensed_galaxies_HATLAS_vs_HeLMSplusHerS}
\end{figure}

\begin{figure}
  \hspace{-0.5cm}\includegraphics[width=9.0cm]{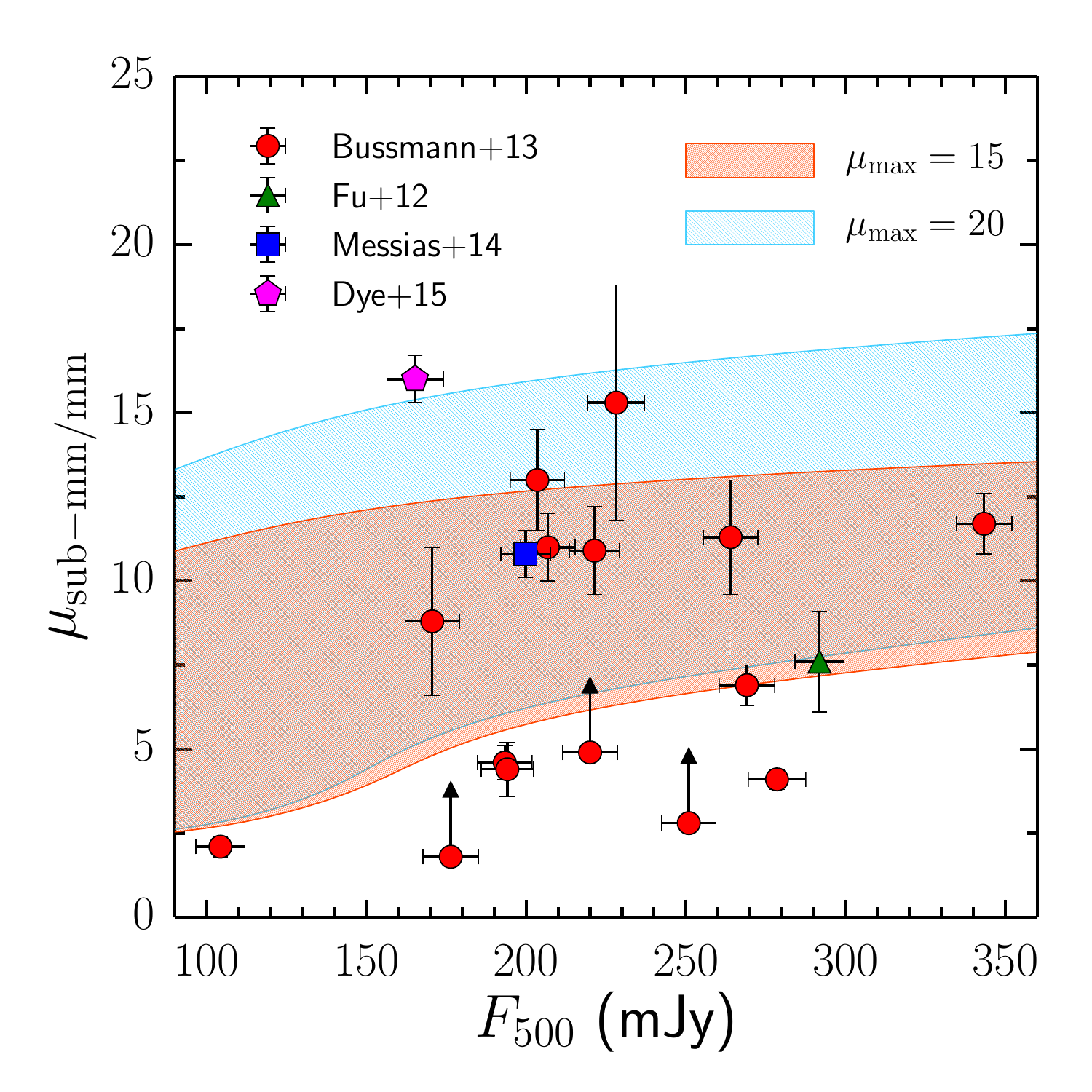}
 \vspace{-0.5cm}
 \caption{Magnification factors derived from high resolution sub-mm/mm imaging of the confirmed lensed galaxies in the GAMA and NGP fields,
compared with model prodictions (Lapi et al. 2012 and Cai et al. 2013) for $\mu_{\rm max}=15$ and $\mu_{\rm max}=20$. The shaded region shows the 
68 per cent confidence interval around the median value of the
probablity distribution of the magnifications. The range of
  maximum magnifications suggested by the data is consistent with that
inferred from the number counts of candidate lensed galaxies with
$F_{\rm 500}\gsim150\,$mJy in the
GAMA and NGP fields (see fig.\,\ref{fig:counts_lensed_galaxies_HATLAS}).} 
 \label{fig:mu_vs_F500_mumax15}
\end{figure}

The function $p(\mu,z_{\rm S})$ depends on several factors: the distribution in mass and in 
redshift of the lenses and their mass profile. For a given lens of mass $M_{\rm L}$ and redshift 
$z_{\rm L}$, and a given redshift of source $z_{\rm S}$, one can define, in the source plane, the 
cross-section, $\Sigma(>\mu|M_{\rm L},z_{\rm L},z_{\rm S})$, for lensing events with total magnification 
$>\mu$, i.e. the area containing all the source positions for which $\mu_{\rm tot}>\mu$. In the strong 
lensing regime the probability $P(>\mu|z_{\rm S})$ that a source is magnified by a value higher than $\mu$ 
is significantly lower than unity (as multiple images are only formed when the source and the lens are sufficiently 
well aligned), and it is very unlikely that the source is lensed by more than one foreground mass
 (i.e. ``non-overlapping'' cross-sections approximation). In this regime the probability $P(>\mu|z_{\rm S})$ 
is just \citep[e.g.][]{P82}
\begin{eqnarray}
P(>\mu|z_{\rm S}) = \nonumber ~~~~~~~~~~~~~~~~~~~~~~~~~~~~~~~~~~~~~~~~~~~~~~~~ \\
\int_{z_{\rm L}^{\rm min}}^{z_{\rm L}^{\rm max}} \int_{M_{\rm L}^{\rm min}}^{M_{\rm L}^{\rm max}} \Sigma(>\mu|M_{\rm L},z_{\rm L},z_{\rm S}) \frac{dN_{\rm L}}{dM_{\rm L}dz_{L}}dM_{\rm L}dz_{\rm L},
\label{eq:Pmu_general_formula}
\end{eqnarray}
where $dN_{\rm L}/dM_{\rm L}dz_{L}$ is the number density of the
lenses per unit interval in mass and in redshift. The (differential) probability distribution of the magnifications can then be obtained by differentiating, i.e. $p(\mu|z_{\rm S})=-dP(>\mu|z_{\rm S})/d\mu$.
The total mass distribution of the lenses is described by the mass function of the dark matter halos derived from N-body simulation \citep[e.g.][]{ST99}. 
Therefore, the main uncertainty in the modelling of $p$ is related to the choice of the
cross-section $\Sigma(>\mu|M_{\rm L},z_{\rm L},z_{\rm S}))$ which
strictly depends on the mass profile of the lenses. A Singular
Isothermal Ellipsoid \citep[SIE;][]{Kor94} model
provides a good description of galaxy-scale lensing events,
i.e. events where the lensing mass is associated with one single galaxy,
typically an elliptical \citep[][and referencees therein]{Barn11}, and
of the corresponding image separation distribution \citep[see
e.g. Fig.\,10 of][]{More12}. As shown by e.g. \cite{Lapi12} the SIE
profile results from the
combination of a Navarro-Frank-White (NFW) profile for the dark matter
distribution and a de Vaucouleurs profile for the distribution of baryons in elliptical
galaxies. Here we adopt the NFW plus de Vaucouleurs  model by Lapi et al. to work
out the probability distribution of the magnifications. For the dark matter component we adopt the 
concentration parameter derived by \cite{Prada2012} from numerical simulation, as a function of the dark matter halo mass and redshift.
We assume a dark-matter to baryon mass ratio $M_{DM}/M_{\rm baryons}=30$ and a virialization redshift $z_{\rm vir}=3$ for the lens.
These choices are consistent with observations of early-type galaxies \cite[see ][ and references therein]{Lapi12}. We refer the reader to the Lapi et al. paper for a discussion 
of how the probability distribution of magnifications and the lensed number counts are affected by different choices of these parameters. 
For the number counts of unlensed DSFGs we assume the model of
\cite{Cai13} which accurately fits a broad variety of data: multifrequency and multi-epoch luminosity functions of galaxies and AGN, redshift distributions, number counts (total and per redshift bins).
The derived model predictions for different values of the maximum magnification,
in the range 10 to 30, are shown in
fig.\,\ref{fig:counts_lensed_galaxies_HATLAS} and in fig.\,\ref{fig:counts_lensed_galaxies_HATLAS_vs_HeLMSplusHerS}.
The number counts of the whole sample of candidate lensed galaxies
(Fig.\,\ref{fig:counts_lensed_galaxies_HATLAS_vs_HeLMSplusHerS}), as
well as the redshift distribution (Fig.\,\ref{fig:Nz_lensed_galaxies}),
suggest $\mu_{\rm max}\sim$10$-$15, although in the GAMA\,9h and
  NGP fields higher
  values of the maximum magnification, i.e. 
  $\mu_{\rm max}\gsim$15$-$20, are in better agreement with the 
  observations for $F_{\rm
  500}\gsim150\,$mJy. As shown in
  fig.\,\ref{fig:mu_vs_F500_mumax15} these values of $\mu_{\rm max}$ are consistent with
the magnification factors derived from the modeling of high resolution sub-mm/mm imaging data
of sub-mm selected lensed galaxies
\citep[e.g.][]{Swin10,Buss13,Buss15,Dye15,Spilker16}. 

Using our theoretical models, we have investigated the level of fluctuations  
observed in the number density of candidate lensed
galaxies between the GAMA+NGP and SGP fields. To do so, we have run a set of 100
simulations, over an area of 300\,deg$^{2}$, by re-sampling at random the model luminosity function of
DSFGs and by assigning to each simulated object a probability of being
strongly lensed, and its corresponding magnification factor, according
to the theoretical probability distribution of magnifications. We have
assumed $\mu_{\rm max}=12.5$, the mean value of the
maximum magnifications suggested by the data, i.e. 10 to 15
(see fig.\,\ref{fig:counts_lensed_galaxies_HATLAS_vs_HeLMSplusHerS}). The range of integral number densities of 
lensed galaxies produced by the simulations is shown by the red shaded region in the
bottom-right panel of
Fig.\,\ref{fig:counts_lensed_galaxies_HATLAS}. Large
fluctuations in the number of the brightest lensed galaxes, as those
observed in the GAMA+NGP vs SGP field, are quite rare, but not
unlikely to occurr. In
fact, in
two simulated catalogues (i.e. 2\% of the total) only one lensed source is detected
above $200\,$mJy, while three simulated samples 
(i.e. 3\% of the simulations) contain at least 12
sources\footnote{12 is the expectation value based on the observed
  number of lensed galaxies in the GAMA+NGP field rescaled to the
  300\,deg$^{2}$ area of the simulations.} with
$F_{\rm 500}\geq200\,$mJy. Overall, we find a good agreement
between the results of the simulations and the observations.

\section{Conclusions}\label{sec:conclusions}

We have presented a catalogue of  80 candidate lensed galaxies with
$F_{\rm 500}\ge100\,$mJy selected from the $\sim$600\,deg$^{2}$ of the full
{\it H}-ATLAS. 20 of them are confirmed to be lensed systems based on
available optical to sub-mm/mm spectroscopic and high resolution
imaging data. One source was found to be a binary system of
HyLIRGs. Other 8 sources are very
likely to be lensed based on the presence of two distinct objects
along the same
line of sight as suggested by the estimated photometric redshifts. 
The remaining candidate lensed galaxies still await follow-up observations to confirm their
nature. 

The number density of our candidate lensed galaxies is in agreement with
expectations based on a physical model for the un-lensed DSFG
population coupled
with a NFW plus de Vaucouleurs model for the mass profile of the lenses,
assuming a maximum magnification in the range 10 to 20. These values
of $\mu_{\rm max}$ are in agreement
with the magnification factors estimated from the lens modeling and source reconstruction of
individual lensed DSFGs. We observe quite a large fluctuation in the number density of very
bright, i.e. $F_{\rm 500}\gsim200\,$mJy, lensed galaxies between the
GAMA+NGP fields (13 detections) and the SGP field (only 1
detection). However we have shown, also by means of numerical
simulations, that these variations, although extreme, are still
consistent with expectations based on Poisson statistics.
Once we combine the samples from
all the {\it H}-ATLAS fields, the derived number counts are consistent
with those measured for the candidate lensed galaxies in HeLMS+HerS \citep{Nayyeri16}.

This {\it H}-ATLAS catalogue of candidate lensed galaxies will provide valuable targets for follow-up observations
aimed at studying with unprecedented detail the morphological and dynamical properties
of dusty starforming galaxies at $z\sim2$, as recently demonstrated
by the analysis of the high-resolution ALMA
observations of SDP.81 \citep{ALMA15,Rybak15a,Rybak15b,Dye15,Swin15,Tamura15,Hatsukade15,Hezaveh16}, one of the first lensed galaxies
discovered with {\it Herschel} \citep{Neg10}. 

$~$ \\
{\bf Acknowledgments} \\
We thank the anonymous referee for helpful suggestions and useful
comments.
This project has received funding from the European Union's Horizon 2020 research and innovation programme under the Marie Sk{\l}odowska-Curie grant agreement No 707601.
This work was supported by ASI/INAF Agreement 2014-024-R.0, by PRIN-INAF 2012
project ``Looking into the dust-obscured phase of galaxy formation
through cosmic zoom lenses in the Herschel Astrophysical Large Area
Survey''. 
J.G.N. acknowledges financial support from the Spanish MINECO for a
‘Ramon y Cajal’ fellowship (RYC-2013-13256) and 
the I+D 2015 project AYA2015-65887-P (MINECO/FEDER). L.D., R.J.I. and S.J.M. acknowledge support from the European Research Council
Advanced Investigator grant, COSMICISM. L.D. and S.J.M. also acknowledge
support from the European Research Council Consolidator grant,
cosmicdust. S.D. acknowledges support by a STFC Ernest Rutherford
Fellowship. C.F. acknowledges funding from CAPES (proc. 12203-1). J.L.W.
is supported by a European Union 
COFUND/Durham Junior Research Fellowship under EU grant agreement
number 267209, 
and acknowledges additional support from STFC (ST/L00075X/1).
C.E.P, C.T. G.V. and L.V.E.K. are supported through an NWO-VICI grant (project
number 639.043.308).
LM is supported by the Science and Technology Facilities Council grant ST/J001597/1 and by a South African DST-NRF Fellowship for Early Career Researchers from the UK.
Some of the observations reported in this paper were obtained with the Southern African Large Telescope (SALT) 
under proposal 2015-2-MLT-006.
Herschel is an ESA space observatory with science instruments provided
by European-led Principal Investigator consortia and with important
participation from NASA. 
The {\it Herschel}-ATLAS is a project with {\it Herschel}, which is an
ESA space observatory with science instruments provided by
European-led Principal Investigator consortia and with important
participation from NASA. The {\it H}-ATLAS website is
http://www.h-atlas.org/. 
The Submillimeter Array is a joint project between the Smithsonian
Astrophysical Observatory 
and the Academia Sinica Institute of Astronomy and Astrophysics and is
funded by the Smithsonian 
Institution and the Academia Sinica.
Support for CARMA construction was derived from the G. and B. Moore
Foundation, the K. T. and E. L. Norris Foundation, the Associates of
the California Institute of Technology, the states of California,
Illinois, and Maryland, and the NSF. CARMA development and
operations were supported by the NSF under a cooperative agreement, and by the CARMA partner universities.
This research has made use of {\sc ALADIN} and of the NASA/IPAC Extragalactic Database (NED) which 
is operated by the Jet Propulsion Laboratory, California Institute of Technology, 
under contract with the National Aeronautics and Space Administration.


$~$ \\
$~$ \\
{\it
$^{1}$School of Physics and Astronomy, Cardiff University, The Parade, Cardiff CF24 3AA, UK \\  
$^{2}$Department of Physical Sciences, The Open University, Walton Hall, Milton Keynes MK7 6AA, UK \\
$^{3}$CAS Key Laboratory for Research in Galaxies and Cosmology, Department of Astronomy, University of Science and Technology of China, Hefei, Anhui 230026, China \\
$^{4}$Dipartimento di Fisica, Universita Tor Vergata, Via Ricerca Scientifica 1, 00133 Roma, Italy \\
$^{5}$SISSA, Via Bonomea 265, I-34136 Trieste, Italy \\
$^{6}$Departamento de F\'{i}sica, Universidad de Oviedo, C. Calvo Sotelo s/n, 33007 Oviedo, Spain \\
$^{7}$INAF, Osservatorio Astronomico di Padova, Vicolo Osservatorio 5, I-35122 Padova, Italy \\
$^{8}$School of Physics and Astronomy, University of Nottingham, University Park, Nottingham NG7 2RD, UK \\
$^{9}$CAPES Foundation, Ministry of  Education of Brazil, Bras\'ilia/DF, 70040-020, Brazil \\
$^{10}$Institute for Astronomy, University of Edinburgh, Royal Observatory, Blackford Hill, Edinburgh EH9 3HJ, UK \\
$^{11}$Department of Astronomy, Space Science Building, Cornell University, Ithaca, NY 14853, USA \\
$^{12}$Dept. of Physics \& Astronomy, University of California, Irvine, CA 92697, USA \\
$^{13}$Dipartimento di Fisica, Università di Napoli "Federico II", Via Cinthia, I-80126 Napoli, Italy \\
$^{14}$Instituto de Astrofısica de Canarias, E-38205 La Laguna, Tenerife, Spain \\
$^{15}$Department of Physics \& Astronomy, University of Iowa, Iowa
City, Iowa 52242 \\
$^{16}$Imperial College London, Blackett Laboratory, Prince Consort Road, London SW7 2AZ, UK \\
$^{17}$Harvard-Smithsonian Center for Astrophysics, 60 Garden Street, Cambridge, MA 02138, USA \\
$^{18}$Kapteyn Astronomical Institute, University of Groningen, PO Box 800, NL-9700 AV Groningen, the Netherlands \\
$^{19}$INAF-Osservatorio Astronomico di Capodimonte, Via Moiariello 16, I-80131 Napoli, Italy \\
$^{20}$UPMC Univ. Paris 06, UMR7095, Institut d'Astrophysique de Paris, 75014 Paris, France \\
$^{21}$CNRS, UMR7095, Institut d'Astrophysique de Paris, 75014 Paris, France \\
$^{22}$Centre for Extragalactic Astronomy, Department of Physics, Durham University, South Road, Durham, DH1 3LE, UK \\
$^{23}$Sterrenkundig Observatorium, Universiteit Gent, Krijgslaan 281
S9, B-9000 Gent, Belgium \\ 
$^{24}$Department of Physics and Astronomy, Rutgers, the State
University of New Jersey, 136 Frelinghuysen Road, Piscataway, NJ
08854-8019, USA \\
$^{25}$South African Astronomical Observatory, P.O. Box 9 Observatory 7935, Cape Town, South Africa \\
$^{26}$European Southern Observatory, Karl-Schwarzschild-Strasse 2, 85748 Garching, Germany \\
$^{27}$Department of Physics and Astronomy, University of the Western
Cape, Robert Sobukwe Road, 7535 Bellville, Cape Town, South Africa \\
$^{28}$INAF - Istituto di Radioastronomia, via Gobetti 101, 40129
Bologna, Italy \\
$^{29}$Leiden Observatory, Leiden University, P.O. Box 9513, NL-2300
RA Leiden, The Netherlands 
}

\appendix

\section{Postage stamps of candidate lensed galxies}\label{app:stamps}

We show here optical to near-IR postage stamp
images of the candidate lesed galaxies  with $F_{\rm 500}\ge 100\,$mJy identified
in the {\it H}-ATLAS fields, and with lensing rank B or C.

\setcounter{figure}{0}
\begin{figure*}
\vspace{-0.0cm}
\centering
  \hspace{0.0cm}
\includegraphics[width=16cm]{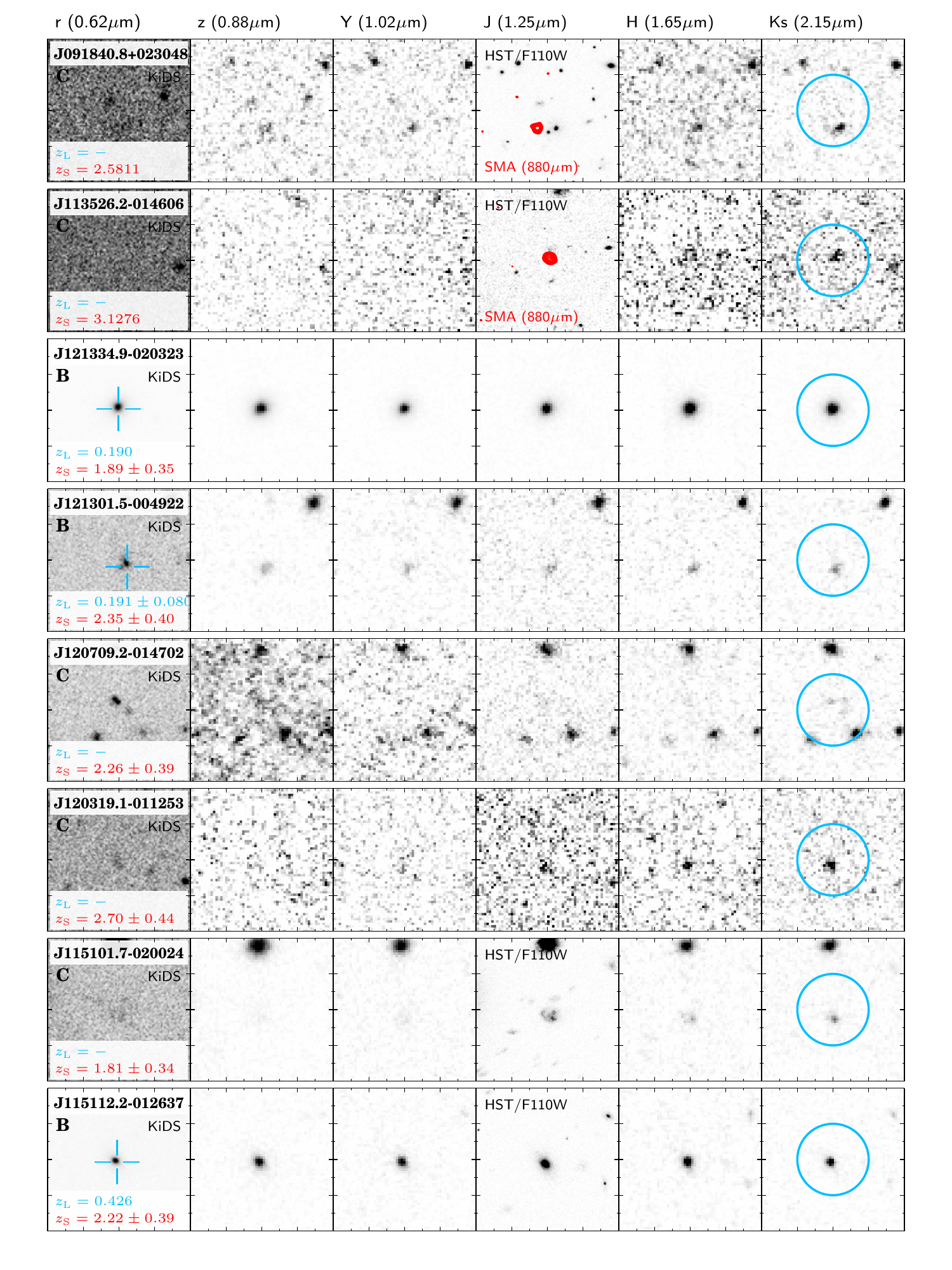}
 \vspace{-0.5cm}
 \caption{
$20^{\prime\prime}\times20^{\prime\prime}$ postage stamps of
 candidate lensed galaxies with $F_{\rm 500}\ge 100\,$mJy and with lensing rank B or C extracted from the {\it H}-ATLAS equatorial fields, centered at
 the position of the {\it Herschel}/SPIRE detection. Imaging data are from KiDS (r band) and VIKING (Z, Y, J, H, Ks bands). The VIKING J-band image is replaced by the {\it HST}/F110W image when {\it HST} data are available.
 For each source, the optical and the sub-mm redshifts are shown at the bottom of the $r$-band stamp, while, in the same stamp, the lensing rank is shown on the top-left corner. The uncertainty on redshift is reported only when the value
 of $z$ is photometrically derived. Where a reliable ($R>0.8$) optical counterpart is found (see Table\,\ref{tab:lens_candidates_F500ge100mJy}) 
its position is marked by a cross in the $r$-band images. A circle of
$5^{\prime\prime}$ in radius, centered at the position of the SPIRE
detection, is shown in the K$_{s}$ images. Where available SMA contours
at 880\,$\mu$m are shown in red.
}
 \label{fig:stamps_GAMA}
\end{figure*}
\setcounter{figure}{0}
\begin{figure*}
\vspace{0.0cm}
\centering
\hspace{-0.0cm}
\includegraphics[width=16cm]{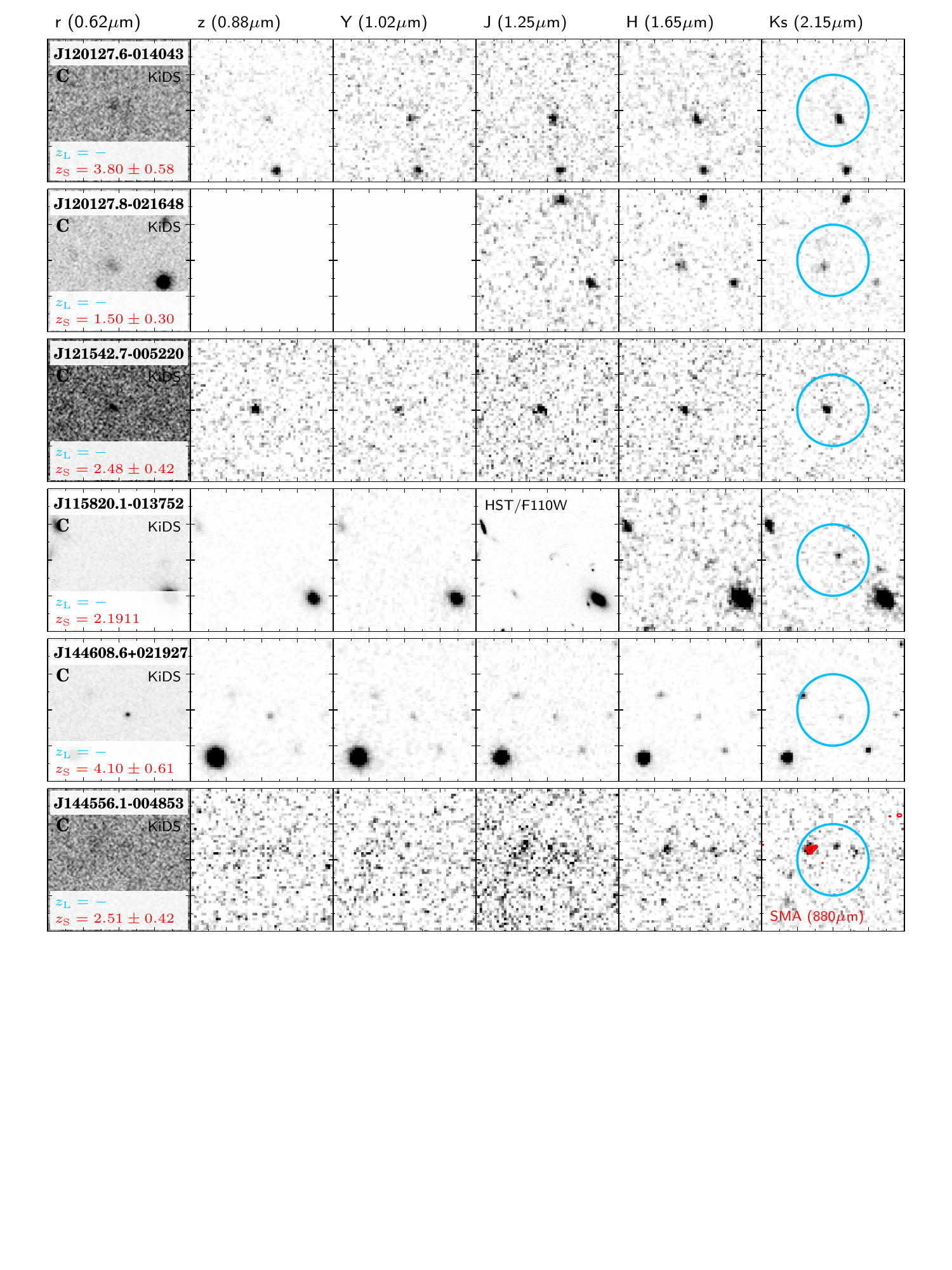}
 \vspace{-5.5cm}
 \caption{\it Continued.}
\end{figure*}

\setcounter{figure}{1}
\begin{figure*}
\vspace{-0.0cm}
\centering
  \hspace{0.0cm}
\includegraphics[width=16cm]{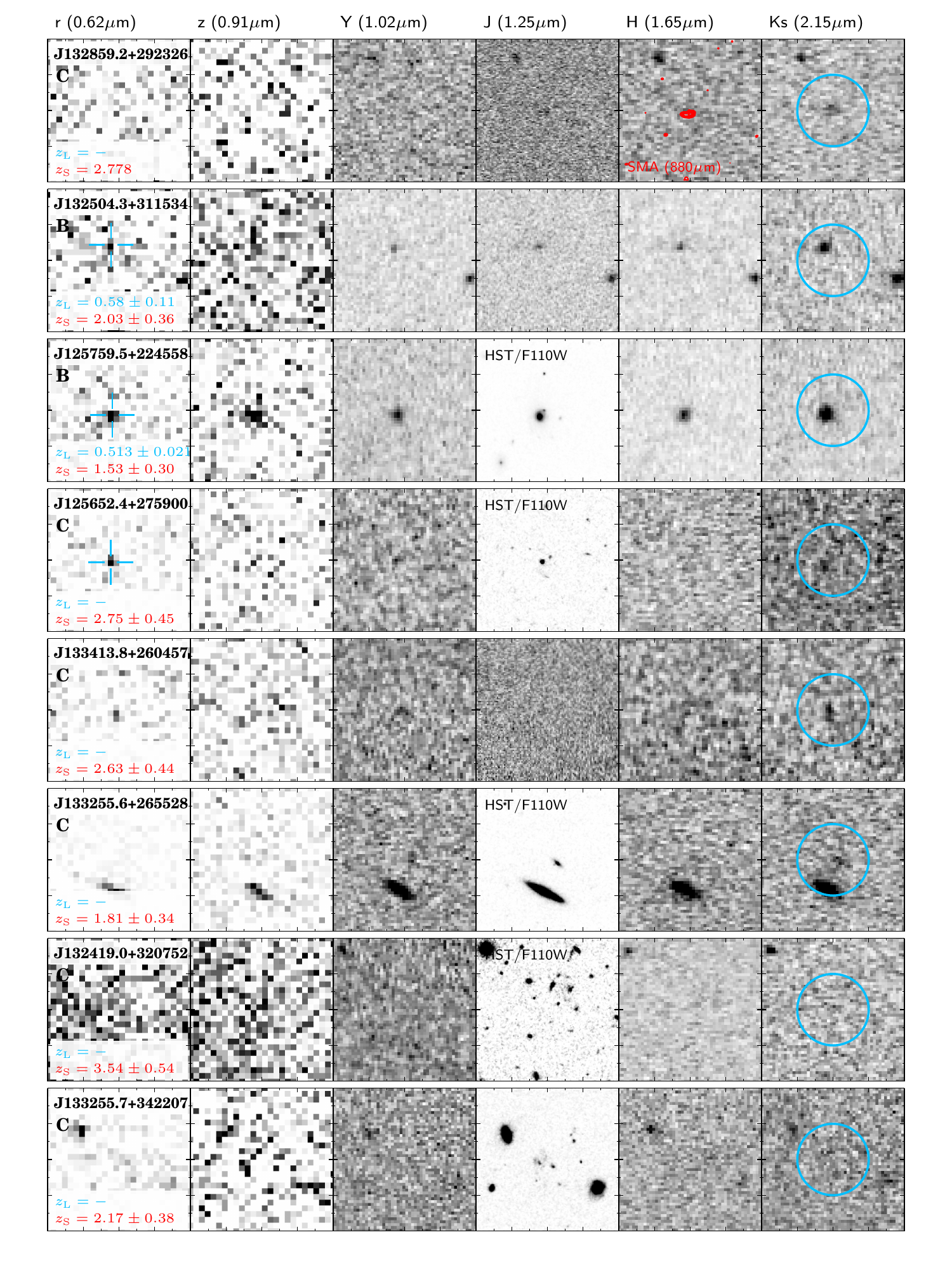}
 \vspace{-0.5cm}
 \caption{
$20^{\prime\prime}\times20^{\prime\prime}$ postage stamps of
 candidate lensed galaxies with $F_{\rm 500}\ge 100\,$mJy and with lensing rank B or C extracted from the {\it H}-ATLAS/NGP field, centered at
 the position of the {\it Herschel}/SPIRE detection. Imaging data are from SDSS (r and z bands) and UKIDSS-LAS (Y, J, H, Ks bands). The UKIDSS J-band image is replaced by the {\it HST}/F110W image when {\it HST} data are available.
 For each source, the optical and the sub-mm redshifts are shown at the bottom of the $r$-band stamp, while, in the same stamp, the lensing rank is shown on the top-left corner. The uncertainty on redshift is reported only when the value
 of $z$ is photometrically derived. Where a reliable optical ($R>0.8$) counterpart is found (see Table\,\ref{tab:lens_candidates_F500ge100mJy}) 
its position is marked by a cross in the $r$-band images. A circle of
$5^{\prime\prime}$ in radius, centered at the position of the SPIRE
detection, is shown in the K$_{s}$ images. Where available SMA contours
at 880\,$\mu$m are shown in red.
}
 \label{fig:stamps_NGP}
\end{figure*}
\setcounter{figure}{1}
\begin{figure*}
\vspace{0.0cm}
\centering
\hspace{-0.0cm}
\includegraphics[width=16cm]{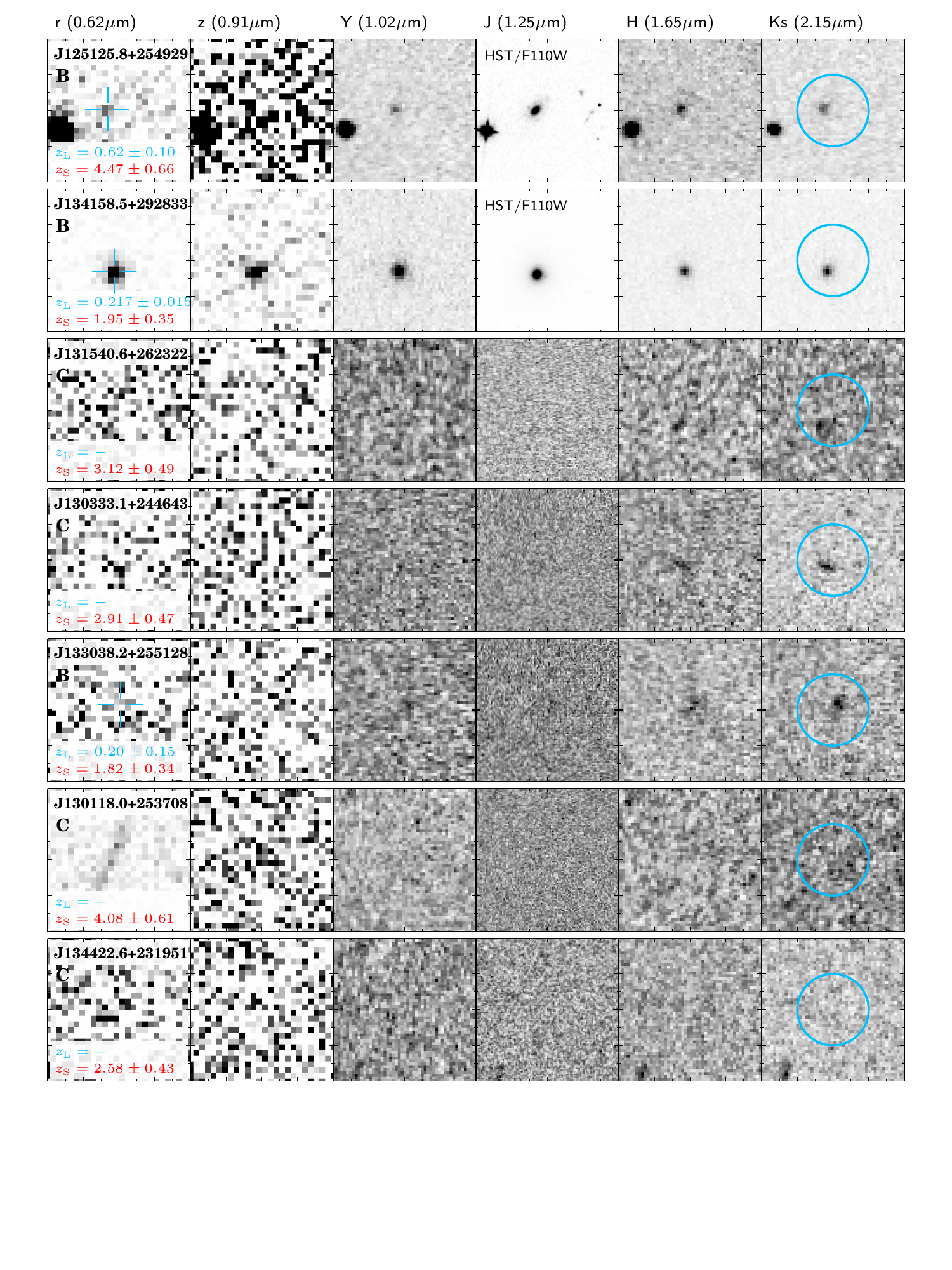}
 \vspace{-0.5cm}
 \caption{\it Continued.}
\end{figure*}

\setcounter{figure}{2}
\begin{figure*}
\vspace{-0.0cm}
\centering
  \hspace{0.0cm}
\includegraphics[width=16cm]{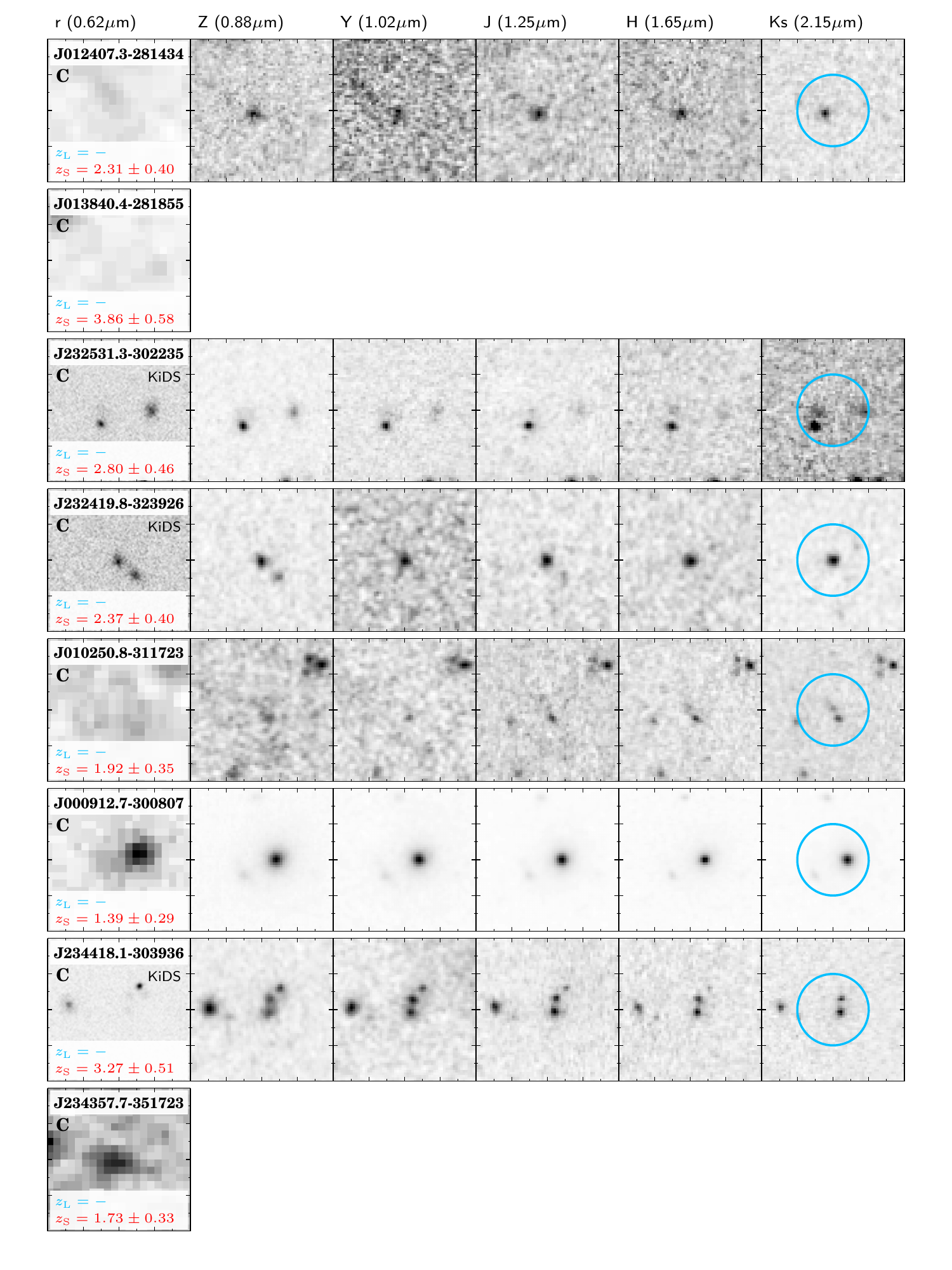}
 \vspace{-0.5cm}
 \caption{$20^{\prime\prime}\times20^{\prime\prime}$ postage stamps of
 candidate lensed galaxies with $F_{\rm 500}\ge 100\,$mJy and with lensing rank B or C extracted from the {\it H}-ATLAS/SGP field, centered at
 the position of the {\it Herschel}/SPIRE detection. Imaging data are
 from the Digitalized Sky Survey (DSS; r band), or KiDS (r band) where available, and VIKING (Z, Y, J, H, Ks bands). 
 For each source, the optical and the sub-mm redshifts are shown at the bottom of the $r$-band stamp, while, in the same stamp, 
the lensing rank is shown on the top-left corner. The uncertainty on redshift is reported only when the value
 of $z$ is photometrically derived. A circle of
$5^{\prime\prime}$ in radius, centered at the position of the SPIRE
detection, is shown in the K$_{s}$ images. }
 \label{fig:stamps_SGP}
\end{figure*}
\setcounter{figure}{2}
\begin{figure*}
\vspace{0.0cm}
\centering
\hspace{-0.0cm}
\includegraphics[width=16cm]{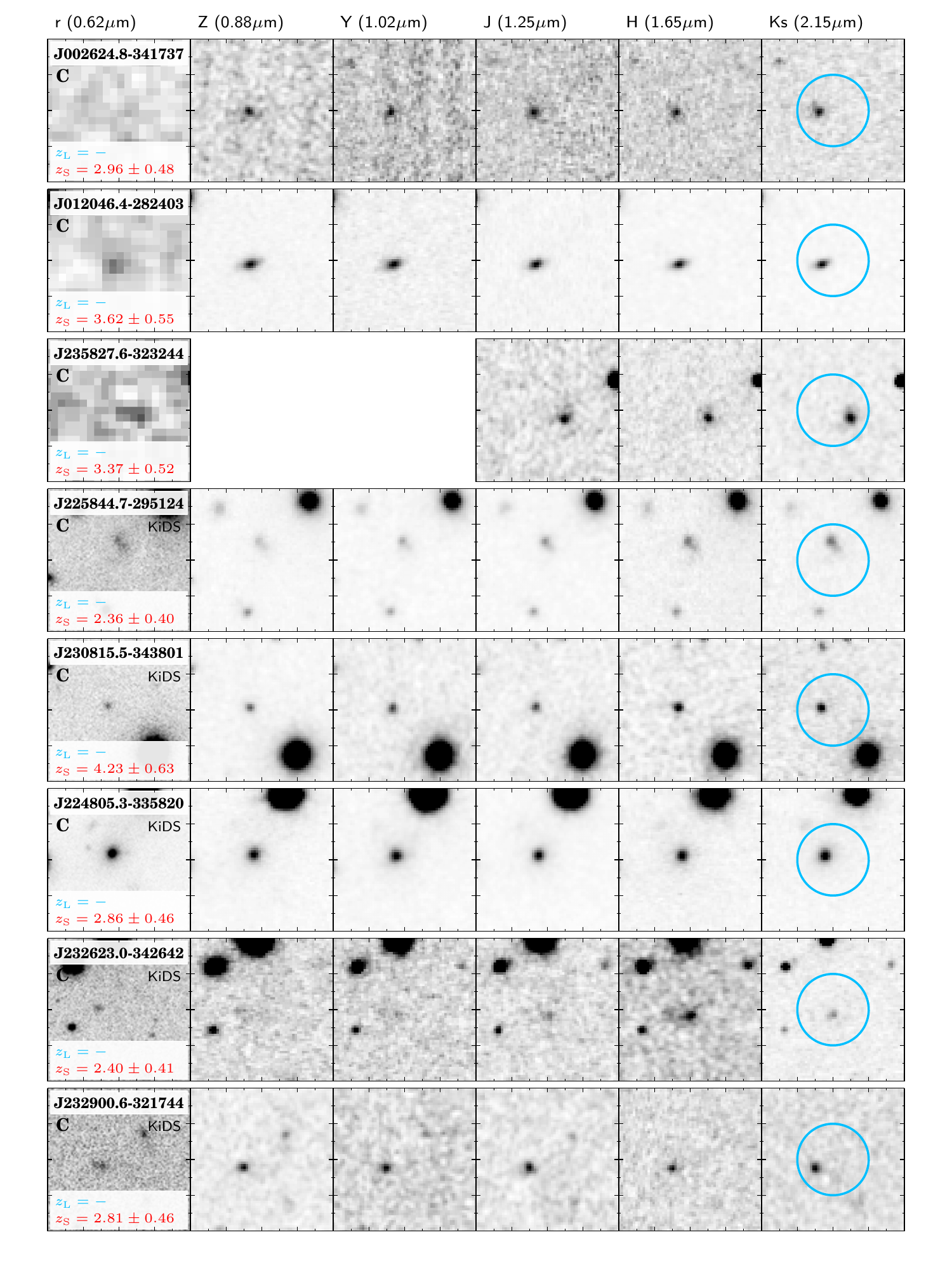}
 \vspace{-0.5cm}
 \caption{\it Continued.}
\end{figure*}
\setcounter{figure}{2}
\begin{figure*}
\vspace{0.0cm}
\centering
\hspace{-0.0cm}
\includegraphics[width=16cm]{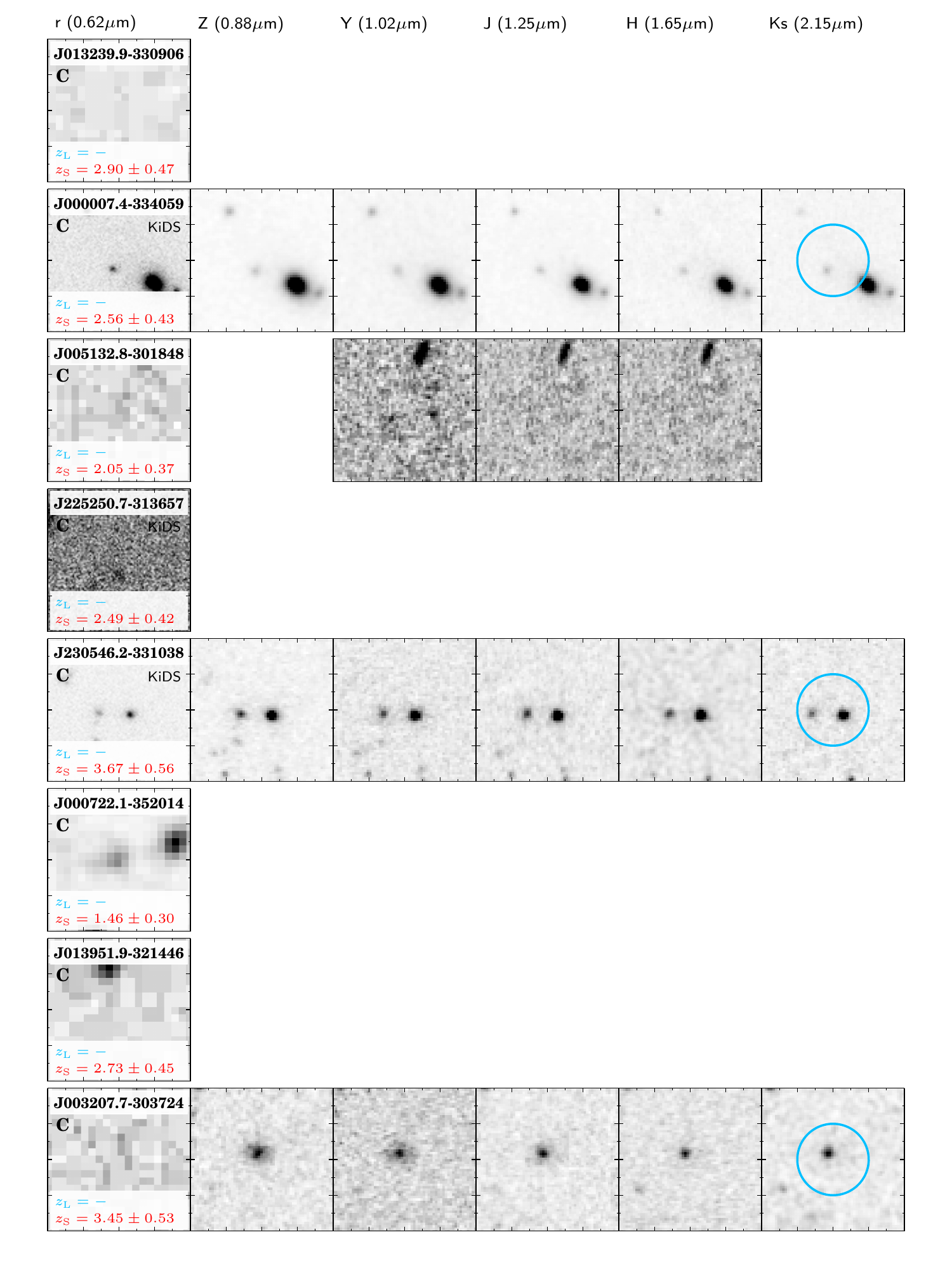}
 \vspace{-0.5cm}
 \caption{\it Continued.}
\end{figure*}
\setcounter{figure}{2}
\begin{figure*}
\vspace{0.0cm}
\centering
\hspace{-0.0cm}
\includegraphics[width=16cm]{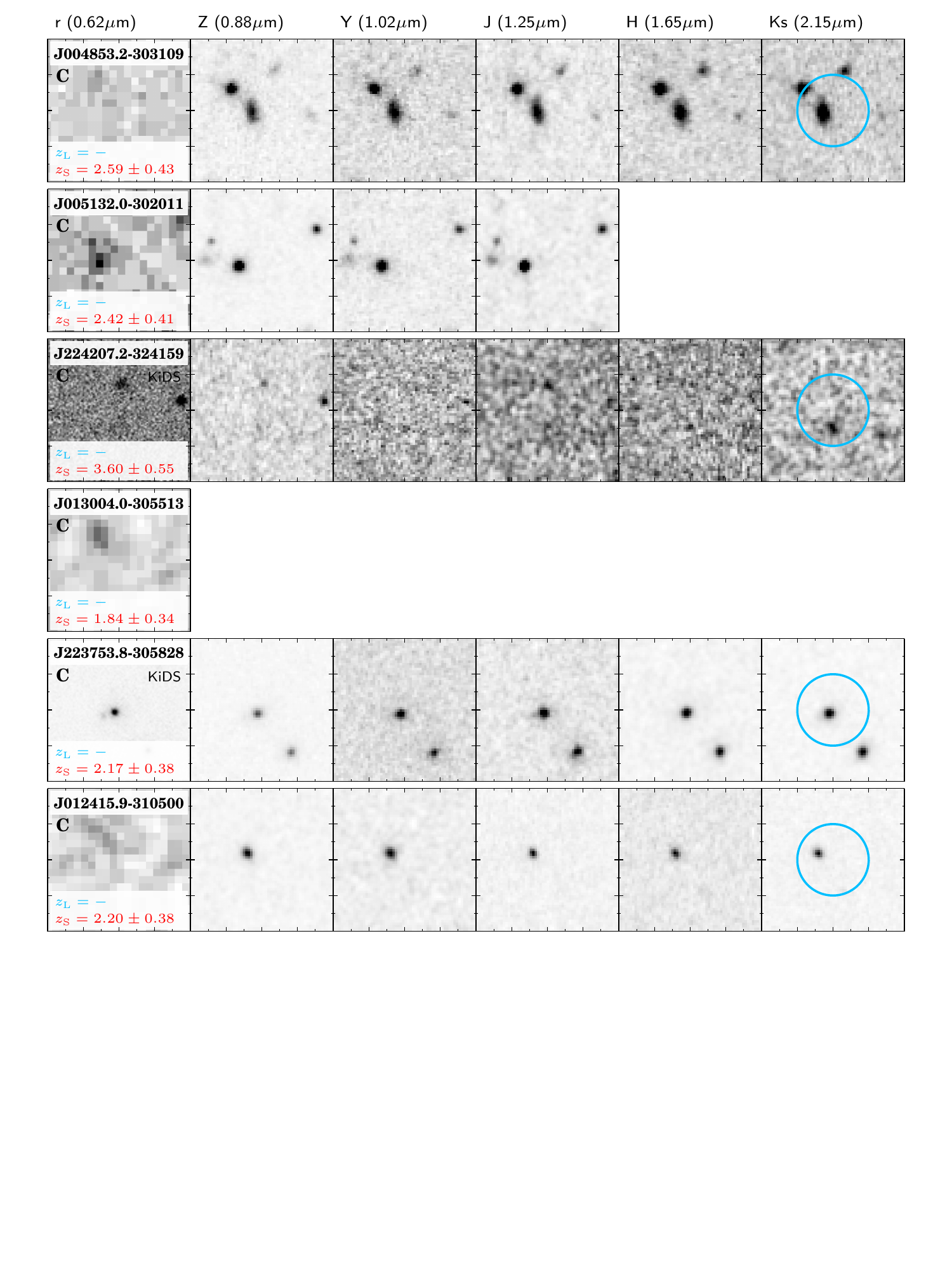}
 \vspace{-5.5cm}
 \caption{\it Continued.}
\end{figure*}

\end{document}